\documentclass[fleqn,usenatbib]{mnras}

\usepackage{newtxtext,newtxmath}

\usepackage[T1]{fontenc}
\usepackage{ae,aecompl}

\usepackage{graphicx}	
\usepackage{amsmath}	
\usepackage{amssymb}	
\usepackage{natbib}
\usepackage{url}
\usepackage{array}
\usepackage{footnote}
\makesavenoteenv{tabular}
\makesavenoteenv{table}
\usepackage{threeparttable}

\def\Teff{\ensuremath{T_\mathrm{eff}}}
\def\Rstar{\ensuremath{R_\mathrm{\star}}}
\def\mpers{\ensuremath{\mathrm{m\,s}^{-1}}}
\def\kmpers{\ensuremath{\mathrm{km\,s}^{-1}}}
\def\fluxcgs{\ensuremath{\mathrm{erg}\,\mathrm{s}^{-1}\,\mathrm{cm}^{-2}}}
\newcommand{\rjup}{\ensuremath{R_\mathrm{Jup}}}
\newcommand{\mjup}{\ensuremath{M_\mathrm{Jup}}}
\newcommand{\Mp}{\ensuremath{M_\mathrm{p}}}
\newcommand{\Rp}{\ensuremath{R_\mathrm{p}}}

\title[Discovery of two transiting hot Jupiters from K2]{EPIC229426032~$b$ and EPIC246067459~$b$: discovery and characterization of two new transiting hot Jupiters from K2}

\author[M. G. Soto et al.]{
		   M.~G. Soto,$^{1}$\thanks{E-mail: maritsoto@ug.uchile.cl}
		   M.~R. D\'iaz,$^{1}$
           J.~S. Jenkins,$^{1,2}$
           F. Rojas,$^{3}$ 
           N. Espinoza,$^{4}$ 
           R. Brahm,$^{3,5}$ 
           \newauthor
           H. Drass,$^{5,6}$ 
           M.~I. Jones,$^{7}$ 
           M. Rabus,$^{3,4}$ 
           J. Hartman,$^{8}$ 
           P. Sarkis,$^{4}$ 
           A. Jord\'an,$^{3,5,4}$
           \newauthor
           R. Lachaume,$^{6,4}$ 
           B. Pantoja,$^{1}$
           M. Vu\v{c}kovi\'{c},$^{9}$
           D.~R.~Ciardi,$^{10}$
		   I. Crossfield,$^{11}$
           \newauthor
           C. Dressing,$^{12}$ 
           E. Gonzales,$^{13}$ 
           L. Hirsch$^{12}$
\\
$^{1}$Departamento de Astronom\'ia, Universidad de Chile, Camino El Observatorio 1515, Las Condes, Santiago, Chile\\
$^2$Centro de Astrof\'isica y Tecnolog\'ias Afines (CATA), Casilla 36-D, Santiago, Chile\\
$^{3}$Instituto de Astrof\'isica, Facultad de F\'isica, Pontificia Universidad Cat\'olica de Chile, Av. Vicu\~{n}a Mackenna 4860, 7820436 Macul,\\Santiago, Chile\\
$^{4}$Max-Planck-Institut f\"{u}r Astronomie, K\"{o}nigstuhl 17, 69117 Heidelberg, Germany\\
$^{5}$Millennium Institute of Astrophysics, Santiago, Chile\\
$^{6}$Center of Astro-Engineering UC, Pontificia Universidad Cat\'olica de Chile, Av. Vicu\~{n}a Mackenna 4860, 7820436 Macul, Santiago, Chile\\
$^{7}$European Southern Observatory, Casilla 19001, Santiago, Chile\\
$^{8}$Princeton University, Department of Astrophysical Sciences, Princeton, NJ\\
$^{9}$Instituto de F\'isica y Astronom\'ia, Universidad de Vapara\'iso, Casilla 5030, Valpara\'iso, Chile\\
$^{10}$Caltech/IPAC-NASA Exoplanet Science Institute Pasadena, CA, USA\\
$^{11}$Department of Physics, Massachusetts Institute of Technology, Cambridge, MA, USA\\
$^{12}$Department of Astronomy, University of California, Berkeley, CA, USA\\
$^{13}$Department of Astronomy and Astrophysics, University of California, Santa Cruz, CA 95064, USA
}

\date{Accepted XXX. Received YYY; in original form ZZZ}

\pubyear{2018}

\begin{document}
\label{firstpage}
\pagerange{\pageref{firstpage}--\pageref{lastpage}}
\maketitle

\begin{abstract}
We report the discovery of two hot Jupiters orbiting the stars EPIC229426032 and EPIC246067459. We used photometric data from Campaign 11 and 12 of the {\it Kepler} K2 Mission and radial velocity data obtained using the HARPS, FEROS, and CORALIE spectrographs. 
  EPIC229426032~$b$ and EPIC246067459~$b$ have masses of $1.60^{+0.11}_{-0.11}$ and $0.86^{+0.13}_{-0.12}\,\mjup$, radii of $1.65^{+0.07}_{-0.08}$ and $1.30^{+0.15}_{-0.14}\,\rjup$, and are orbiting their host stars in 2.18 and 3.20-day orbits, respectively. The large radius of EPIC229426032~$b$ leads us to conclude that this candidate corresponds to a highly inflated hot Jupiter. EPIC2460674559~$b$ has a radius consistent with theoretical models, considering the high incident flux falling on the planet. 
We consider EPIC229426032~$b$ to be a excellent system for follow-up studies, since not only is it very inflated, but it also orbits a relatively bright star ($V = 11.6$).  
\end{abstract}

\begin{keywords}
planets and satellites: detection -- planets and satellites: fundamental parameters -- planets and satellites: gaseous planets
\end{keywords}



\section{Introduction}

Since the detection of the first transiting exoplanet \citep[HD 209458~b,][]{charbonneau2000}, the anomalously large radii of many hot Jupiters have been puzzling astronomers trying to understand the formation and composition of these systems. Inflated giant planets have radii larger than what theoretical models predict for their masses \citep{burrows2007,Fortney2007}, and are often found orbiting their host stars at short periods. This has led many groups to link planetary inflation with several effects, most importantly derived from their stellar insolation \citep[for a review of these theories, see][]{Weiss2013}, and based on observational evidence, an insolation limit of $F > 2 \times 10^8\,\fluxcgs$ has been set which can trigger the expansion of the planet \citep{miller2011,Demory2011}. 

With the launch of the NASA {\it Kepler} space mission \citep{Kepler}, later renamed {\it Kepler K2} due to the failure of one of its reaction wheels \citep{K2}, the number of exoplanets detected has witnessed an exponential growth. 
Because ultracool dwarfs and gas giant planet more or less share a common radius, dynamical mass measurements are required to determine whether a transit signal originates from a planet or an ultracool dwarf.
For single-planet systems, this is possible through the radial velocity method, which also provides the high resolution spectra required for the characterization of the host star and, in consequence, the planet.

Currently, researchers working in Chilean institutions have privileged access to state of the art instrumentation for follow-up observation of planetary candidates through radial velocity. This leaded us to create a Chilean-based K2 project (K2CL), focused on the task of selection of planetary candidates through photometry from the K2 mission, and later follow up using high resolution spectrograph. Exciting results have already been published since the project was started \citep[see][]{Espinoza2016,Brahm2016,Jones2017, Brahm2018}.

In this work we report the discovery of two hot Jupiters, orbiting two dwarf stars that represent two different cases of the hot Jupiter-type planets. EPIC 229426032 is an 11.6 magnitude F star visible from the southern hemisphere (Table~\ref{tab:properties}). It was observed during Campaign 11 of the K2 mission, and the planet was found to have a mass of $1.60^{+0.11}_{-0.11}\,\mjup$, but a radius of $1.65^{+0.07}_{-0.08}\,\rjup$, making it a highly inflated hot Jupiter. The next planet, EPIC 246067459~$b$, was found using data from Campaign 12 of K2 to be orbiting a G type star. For this planet, we found a mass of $0.86^{+0.13}_{-0.12}\,\mjup$, and radius of $1.30^{+0.15}_{-0.14}\,\rjup$. Even though the planet is in the hot Jupiter regime and receives a flux above the inflation threshold, it does not show inflation characteristics.

This paper is organized as follows, in Section \ref{sec:data} we present the data obtained for each star, including photometric and spectroscopic observations. In Section \ref{sec:analysis} we analyze and derive the atmospheric parameters and obtain estimates for their stellar parameters such as age, mass, metallicity, effective temperature and rotational velocity. We also model both the radial velocity observations and the light curves, and derive the physical characteristics for each planetary system. In Section \ref{sec:discussion} we show the evidences which imply that EPIC229426032~$b$ corresponds to a highly inflated hot Jupiter, while EPIC246067459~$b$ appears to be consistent with a hydrogen/helium dominated planet with some metal content. Finally, in Section \ref{sec:summary}, we present a summary of our findings.

\begin{figure*}
   \centering
   \includegraphics[width=18cm]{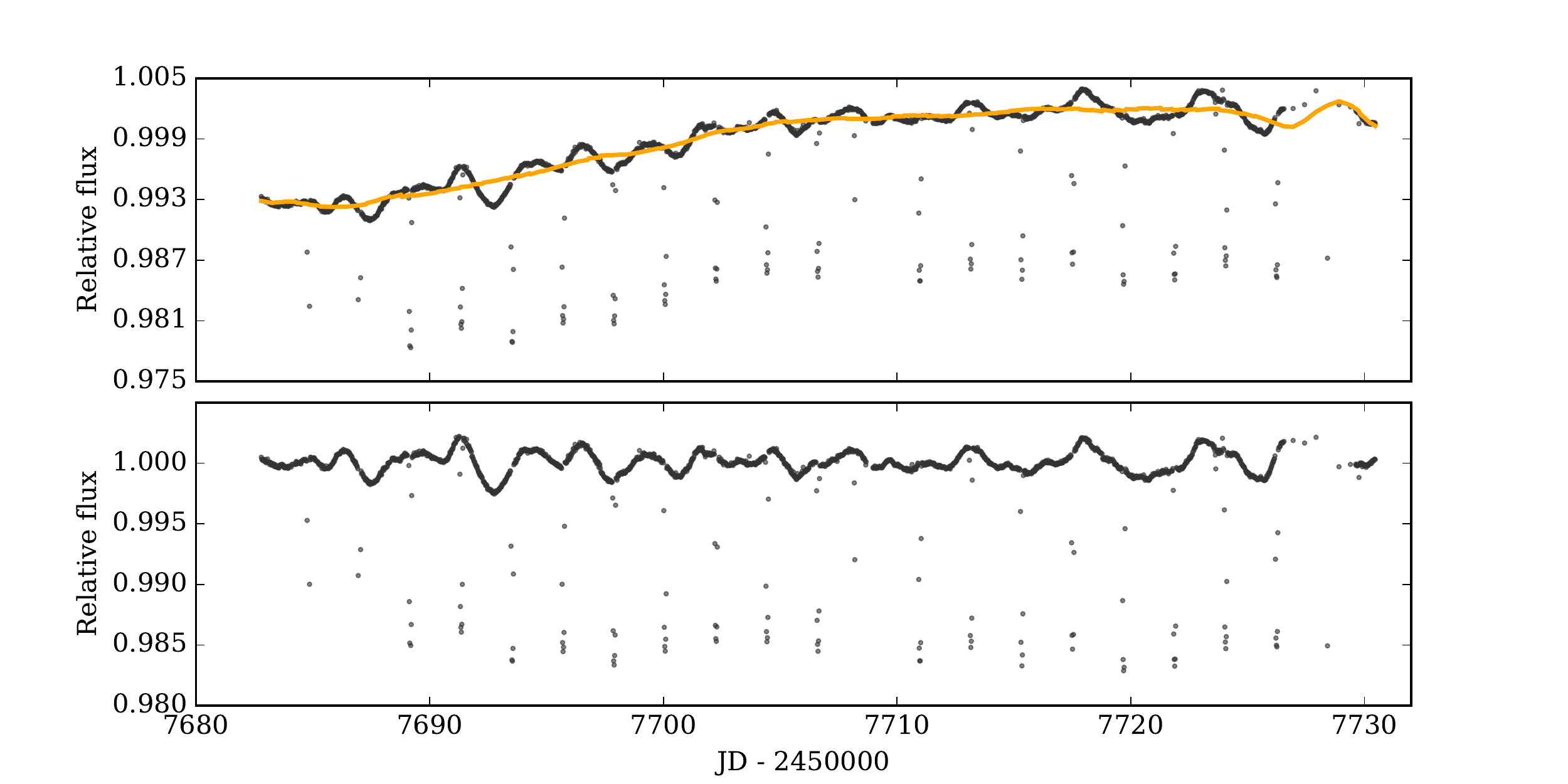}
      \caption{Top panel: Light curve of EPIC229426032 (black), after detrend it with the EVEREST algorithm. The orange line represents the long term variations detected using the polynomial fitting explained in section \ref{sec:photometry}. 
      Bottom panel: Final light curve, with the long term variations removed.}
\label{fig:CL002_lc}
\end{figure*}

\begin{figure*}
\centering
    \includegraphics[width=18cm]{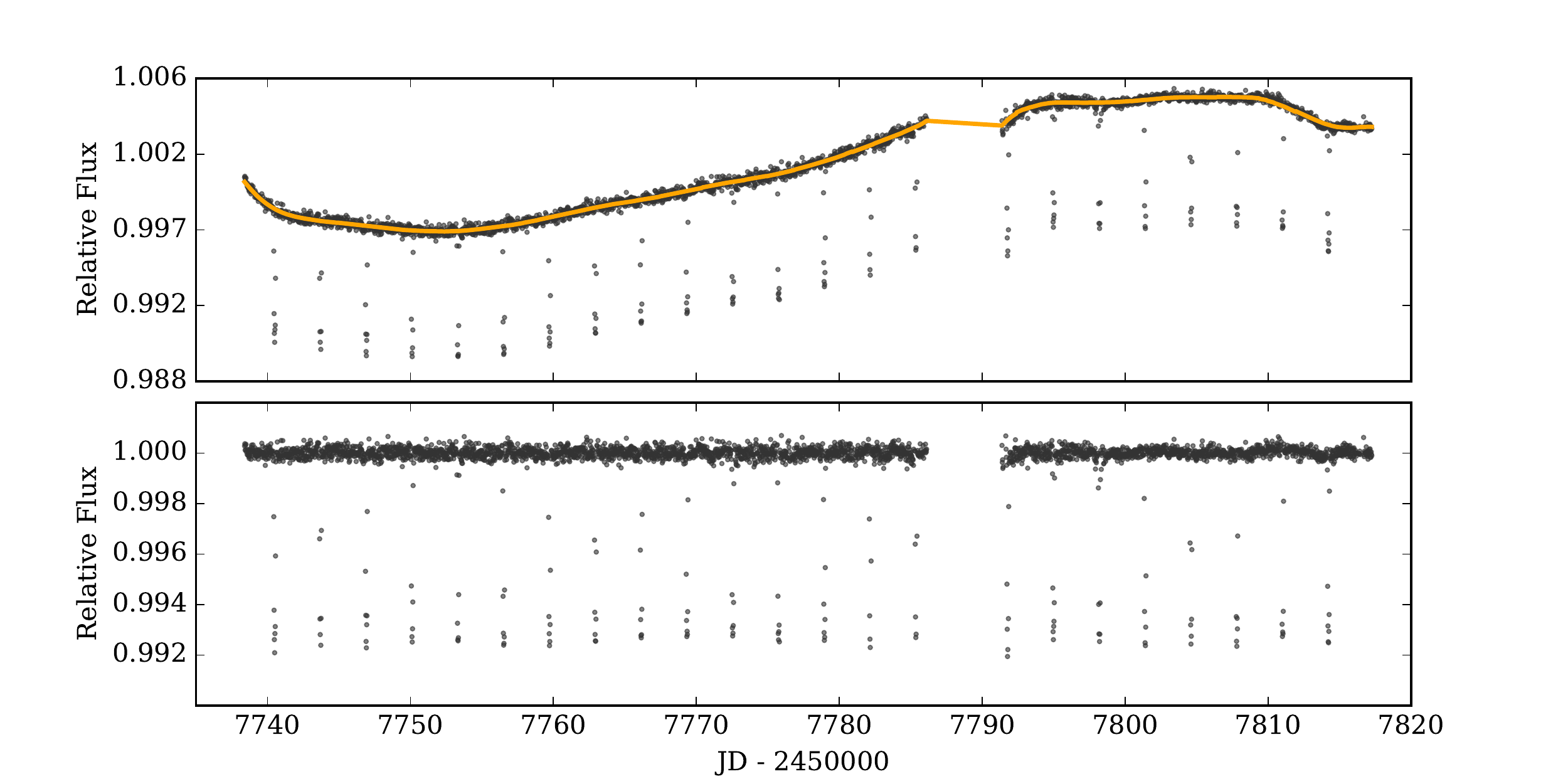}
      \caption{Top panel: Light curve of EPIC246067459 (black), after detrend it with the EVEREST algorithm. The orange line represents the long term variations detected using the polynomial fitting explained in section \ref{sec:photometry}. 
      Bottom panel: Final light curve, with the long term variations removed.}
\label{fig:CL003_lc}
\end{figure*}

\section{Data}\label{sec:data}

\begin{table}
\center
\caption{Stellar Parameters for both stars.}
\label{tab:properties}
{\renewcommand{\arraystretch}{1.5}
\begin{tabular}{lcc}
\hline\hline
 & EPIC 229426032 & EPIC 246067459\\
\multicolumn{1}{l}{Parameter}& &\\
\hline
R.A. (J2000)& \hphantom{-}16:55:04.5& \hphantom{-}23:10:49.042\\
Dec. (J2000)& -28:42:38\hphantom{.0} & -07:51:27.00\hphantom{0}\\
$B$ & 12.19 $\pm$ 0.07 & 14.61 $\pm$ 0.10\\
$V$ & 11.60 $\pm$ 0.05 & 13.75 $\pm$ 0.02\\
$J$ & 10.51 $\pm$ 0.02 & 12.46 $\pm$ 0.03\\
$H$ & 10.27 $\pm$ 0.02 & 12.10 $\pm$ 0.04\\
$K$ & 10.22 $\pm$ 0.02 & 12.03 $\pm$ 0.03\\
Distance (pc) & $458_{-118}^{+196}$ & $453_{-46}^{+72}$\\ 
\hline
Spectral type & F6V & G2V\\
Mass ($M_{\odot}$) &  $1.28_{-0.04}^{+0.03}$ & $1.19_{-0.08}^{+0.08}$\\
Radius ($R_{\odot}$) & $1.43_{-0.07}^{+0.06}$ & $1.59_{-0.16}^{+0.16}$\\
Density ($\rho_{\odot}$) & $0.102_{-0.010}^{+0.012}$ & $0.0550_{-0.0002}^{+0.0003}$\\
T$_{\rm eff}$ (K) & 6257 $\pm$ 100 & 5630 $\pm$ 78\\
\lbrack Fe/H\rbrack  &  0.14 $\pm$ 0.05 &  0.34 $\pm$ 0.04\\
$\log g$ (cm\,s$^{-2}$) & 4.24 $\pm$ 0.10 & 4.11 $\pm$ 0.07 \\
Age (Gyr) & $2.55_{-0.44}^{+0.38}$ & $5.63_{-1.97}^{+1.05}$\\
$P_{\text{rot}}$ (days) & 5.07 $\pm$ 0.02 & \\
$v\sin i$ (km\,s$^{-1}$) & 11.76 $\pm$ 0.90 & 3.78 $\pm$ 0.57 \\
\hline\hline
\end{tabular}}\quad
\vspace{0.5cm}
\end{table}

\subsection{Photometry}\label{sec:photometry}

We analyzed photometric data from Campaign 11 (EPIC229426032) and Campaign 12 (EPIC246067459) of the {\it K2} mission. 
We downloaded the Target Pixel Files (TPF) from MAST, extracted the photometry, and detrended it with an implementation of the EVEREST algorithm \citep{Luger2017}.
The remaining long-term variations were removed following a similar procedure than the one described in \citet{Giles2018}. We locally fit a third-order polynomial to sections of 0.5 days of the light curve, using a window of 10 days over the surrounding data. We repeat this process over the whole light curve. An outlier rejection was performed before fitting the data, to ensure that the transit was not removed. 
The light curves obtained after detrending and removing the long term variations are shown in Figs.~ \ref{fig:CL002_lc}~\&~\ref{fig:CL003_lc}. For the case of EPIC229426032, this is not the final light curve we used to derive the planet parameters. The data we used for that analysis is shown in Figure~\ref{fig:GP_lc}, and the process we followed to process it is explained in section~\ref{sec:rot_period}.

\subsection{Radial Velocity follow-up}

Radial velocity follow-up data for EPIC229426032 was acquired using the CORALIE spectrograph \citep{Queloz2000}, mounted on the 1.2m Euler Swiss Telescope at La Silla Observatory. 

We obtained 9 observations between July 7th and July 11th 2017. For each one of the 4 consecutive nights, we acquired two observations of 1800 seconds each, achieving a signal-to-noise (S/N) ratio of $\sim 20$. The spectra were reduced and analyzed using the CERES automated pipeline \citep{Brahm2017}. The mean radial velocity uncertainty achieved for this target was $\sim 38\,\mpers$. The obtained radial velocities for each epoch are listed in Table~\ref{tab:rvs_coralie}.

We also acquired 4 additional radial velocity data points using HARPS \citep{mayor2003}, which is mounted on the ESO 3.6m telescope at La Silla Observatory. The data were taken during four consecutive nights, with one 1800 seconds exposure per night. The S/N achieved for these data is $\sim 32$. The observations were later processed using the CERES pipeline, obtaining an uncertainty in the radial velocities of $\sim 25\,\mpers$. The HARPS velocities are listed in Table~\ref{tab:rvs_harps}.

For EPIC246067459, 6 radial velocity measurements were obtained using FEROS \citep{kaufer1999}, mounted on the 2.2m ESO/MPG Telescope at La Silla Observatory. The data was taken during five nights between November 6th and November 9th 2017, using exposures of 1500 seconds, and achieving S/N$\sim 32$. The CERES automated pipeline was used to reduce and extract the radial velocities. The mean radial velocity uncertainty achieved with FEROS for this target is $16.5\,\mpers$. The velocities are listed in Table~\ref{tab:rvs_feros}.


\subsection{High-resolution AO Imaging}\label{sec:AO_imaging}

\begin{figure}
\centering
\includegraphics[width=9cm]{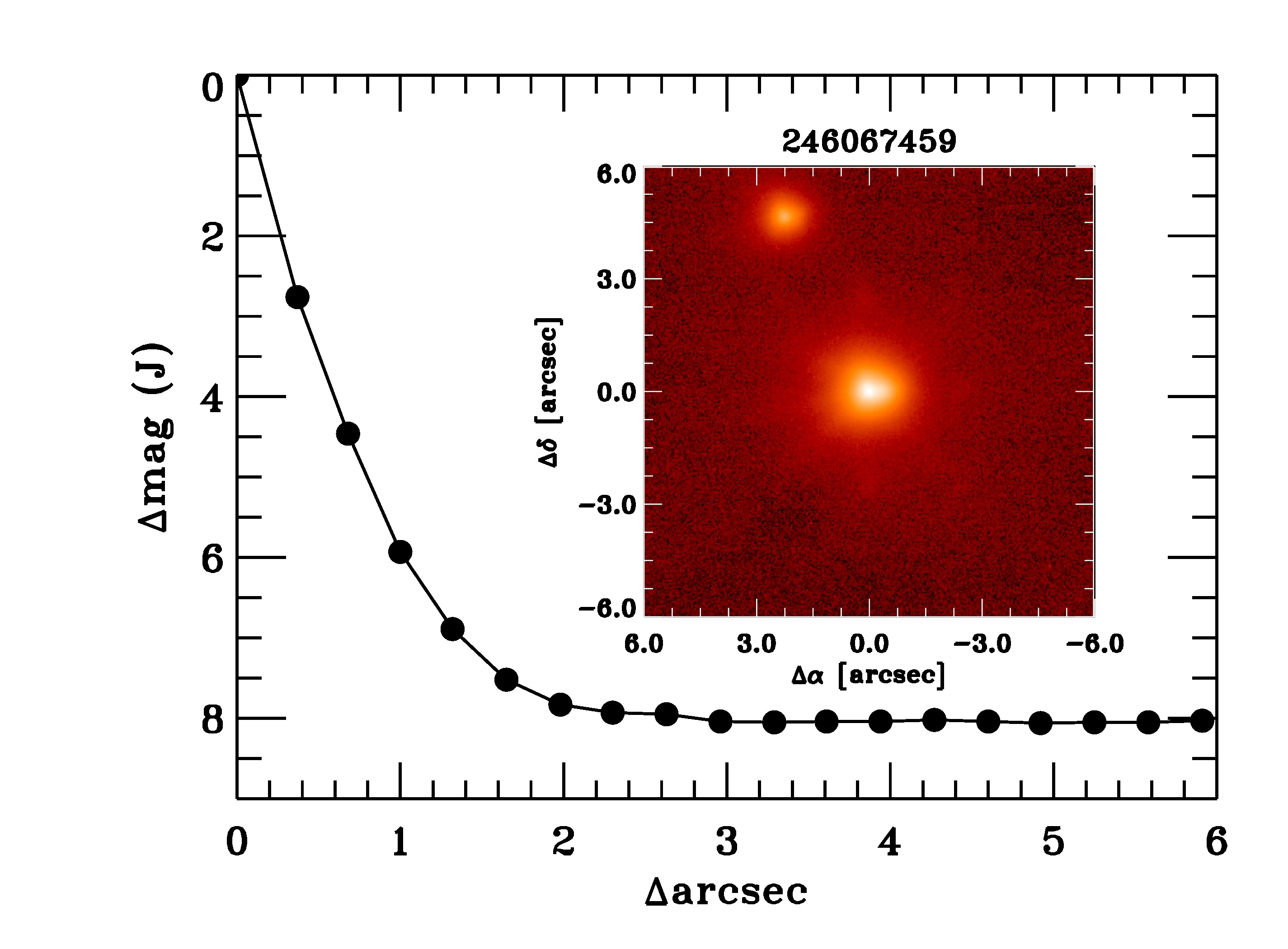}\\
\includegraphics[width=9cm]{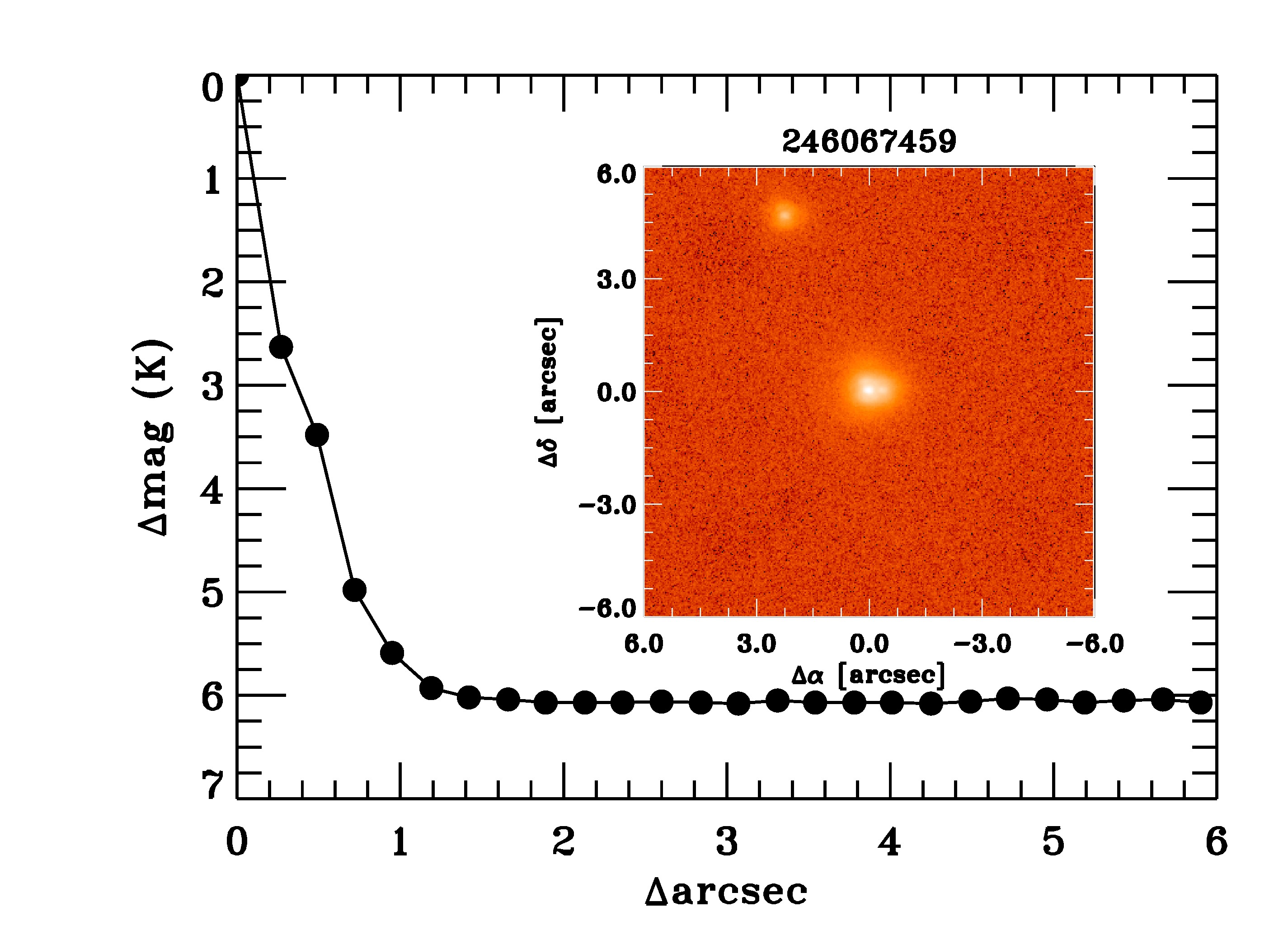}
\caption{Contrast sensitivity curves obtained for EPIC~246067459, in the $J$ (top panel) and $K$ (bottom panel) bands, using ShaneAO at the Lick 3-m telescope. The plot represents the $5\sigma$ contrast limits, in $\Delta$ magnitude, plotted against angular separation in arcseconds. The insets in both figures show the image of the target.}
\label{fig:AO_CL003}
\end{figure}

Observations on the $J$ and $K$ bands for EPIC~246067459 (Figure \ref{fig:AO_CL003}) were taken on August 30, 2017, using the ShaneAO \citep{Gavel2014} at the Lick 3-m Shane Telescope. A PSF of 0.328'' and 0.236''  were obtained for the $J$ and $K$ bands, respectively. The contrast measured at 0.5'' from the center is of $\Delta 2.76$ and $\Delta 3.48$ magnitudes for both bands, respectively. A companion star is seen in both images at around $\sim 2.8''$ from our target (Figure \ref{fig:AO_CL003}).

The photometry was extracted for the resolved companion on both bands, with which we were able to estimate magnitude differences of $\Delta J = 2.2009 \pm 0.0015$ and 
$\Delta K = 2.0009 \pm 0.0053$ with respect to the brighter source, implying $J-K = 0.631 \pm 0.043$. Using this color, we use the \cite{Casagrande2010} color-temperature relations in order to derive a temperature of $T_\textrm{eff} = 4750 \pm 192$ K for the resolved companion, where the error incorporates the uncertainty on the metallicity of the companion (propagated assuming an uniform distribution for it between the validity of the color-temperature relation), the error on our color estimation and the dispersion on the relation itself, which includes uncertainty on the unknown value of $\log(g)$, and which assumes the companion is a dwarf or sub-giant star. 
We could also detect a second companion at 0.35'' from our target. We used aperture photometry to deblend the K-band photometry, obtaining $K=12.47 \pm 0.05$ and $13.2 \pm 0.1$ for the primary star and the companion, respectively. Deblending in the J-band was not possible to perform.

Using the relations from \citet{Howell2012}, we transformed the 2MASS photometry for both stellar companions to the Kepler bandpass, obtaining a magnitude difference with respect to our target of $\Delta K_p = 2.9 \pm 0.8$ and $4.2 \pm 0.6$, for the stars at 0.35'' and 2.8'' away, respectively. We estimate a dilution correcting factor of $1.04 \pm 0.03$ for the radius of the planet orbiting the primary star.


We do not find any close companions to EPIC~229426032 at 5'' from the source.

\section{Analysis}\label{sec:analysis}

\subsection{Stellar Parameters}\label{sec:stellar_params}

The atmospheric parameters for both stars we computed using the Zonal Atmospheric Parameters Estimator \citep[ZASPE,][]{Brahm2017b} code. ZASPE matches the observed stellar spectrum with a set of synthetic spectra generated from the ATLAS9 \citep{ATLAS9} model atmospheres. This procedure is performed via a global $\chi^2$ minimization, in a set of selected spectral regions. For EPIC229426032 we used the co-added CORALIE spectrum, after correcting each individual spectrum by its radial velocity. We used the CORALIE spectra, over the co-added HARPS spectrum, due to the higher S/N obtained. For EPIC246067459 we used the co-added FEROS spectra.

The physical parameters and evolutionary stages of both stars were obtained by interpolating through a grid of Yonsei-Yale isochrones \citep{Demarque2004}. We ran a Markov-Chain Monte Carlo (MCMC), using the \texttt{emcee1} Python package, to explore the parameter space, given by the observed properties of each star. Using the metallicity value derived with ZASPE, we found the posterior distributions for the stellar age and mass. As observed parameters, we use the spectroscopic \Teff{} and the $a/\Rstar$ value obtained from the light curves (see section \ref{sec:joint_analysis}), which is
a more precise proxy for the stellar luminosity than the
spectroscopic $\log(g)$ \citep{sozzetti:2007}.
The derived stellar parameters are listed in Table \ref{tab:properties}.
Both stars have similar masses and are $\approx 25\%$ more massive than the Sun.
While the parameters of EPIC 229426032 are consistent with being in the main sequence, the temperature, radius, and log(g) values of EPIC 246067459 show that it is slightly evolved.
Additionally, both stars, in particular EPIC 246067459 ([Fe/H]=$+0.34$), are enriched in metals compared to the sun.

\subsection{Rotational Period}\label{sec:rot_period}

\begin{figure}
\centering
\includegraphics[width=9cm]{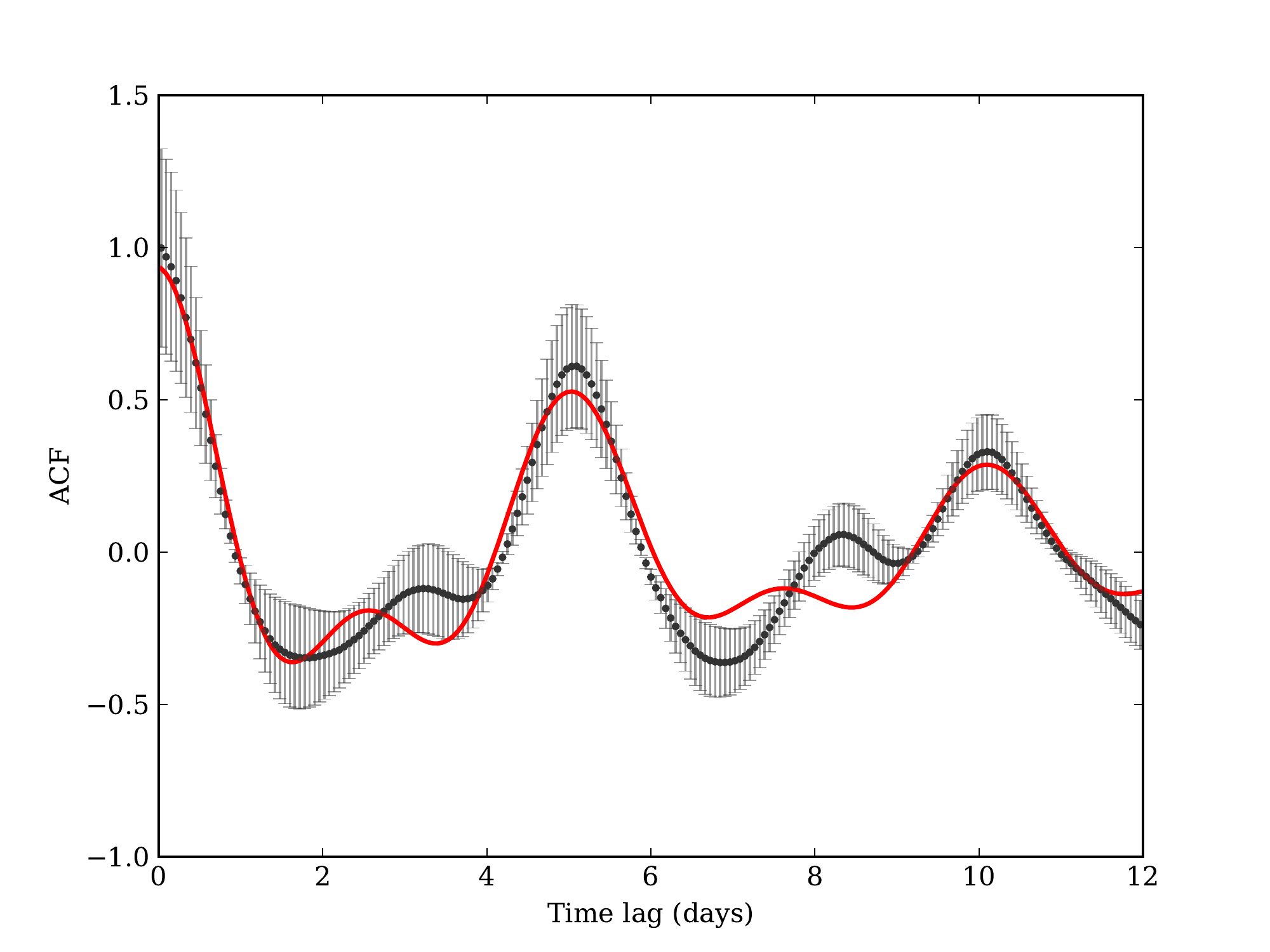}
\caption{Autocorrelation function of the light curve (black dots), along with its uSHO fit (red line), in which the rotation period equals to 5.07 days.}
\label{fig:ACF_odr}
\end{figure}
 
\begin{figure*}
   \centering
   \includegraphics[width=18cm]{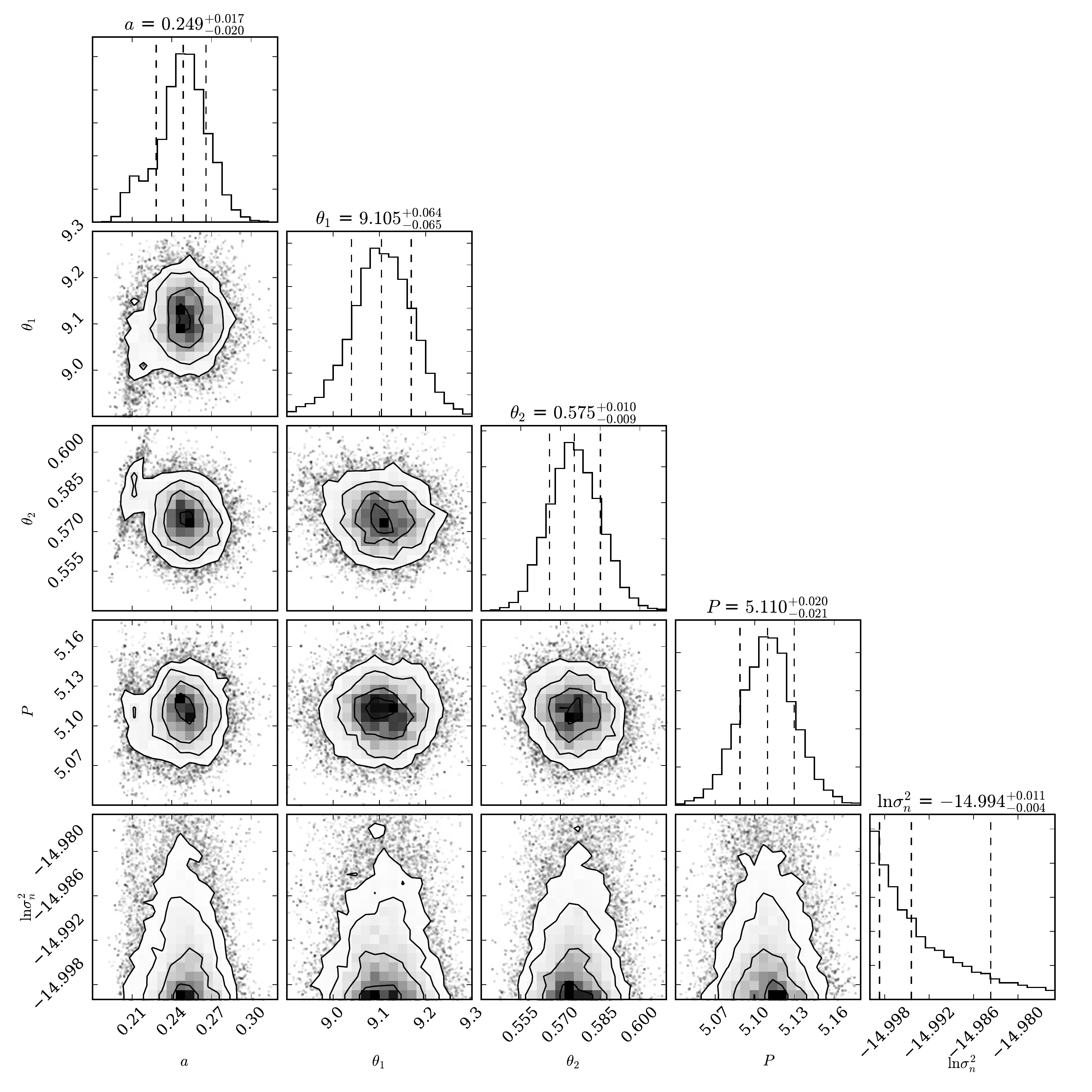}
      \caption{Posterior distributions of the parameters from the Quasi-periodic kernel (Equation \ref{eq:GP}). $a$ is the amplitude of the covariance function, $\theta_1$ is the time scale of the exponential decay, and $\theta_2$ and $P$ are the amplitude and period of the sinusoidal component. $\sigma_n$ corresponds to the amplitude of the white noise component.}
 \label{fig:CL002_GP}
 \end{figure*}

\begin{figure}
\centering
\includegraphics[width=9cm]{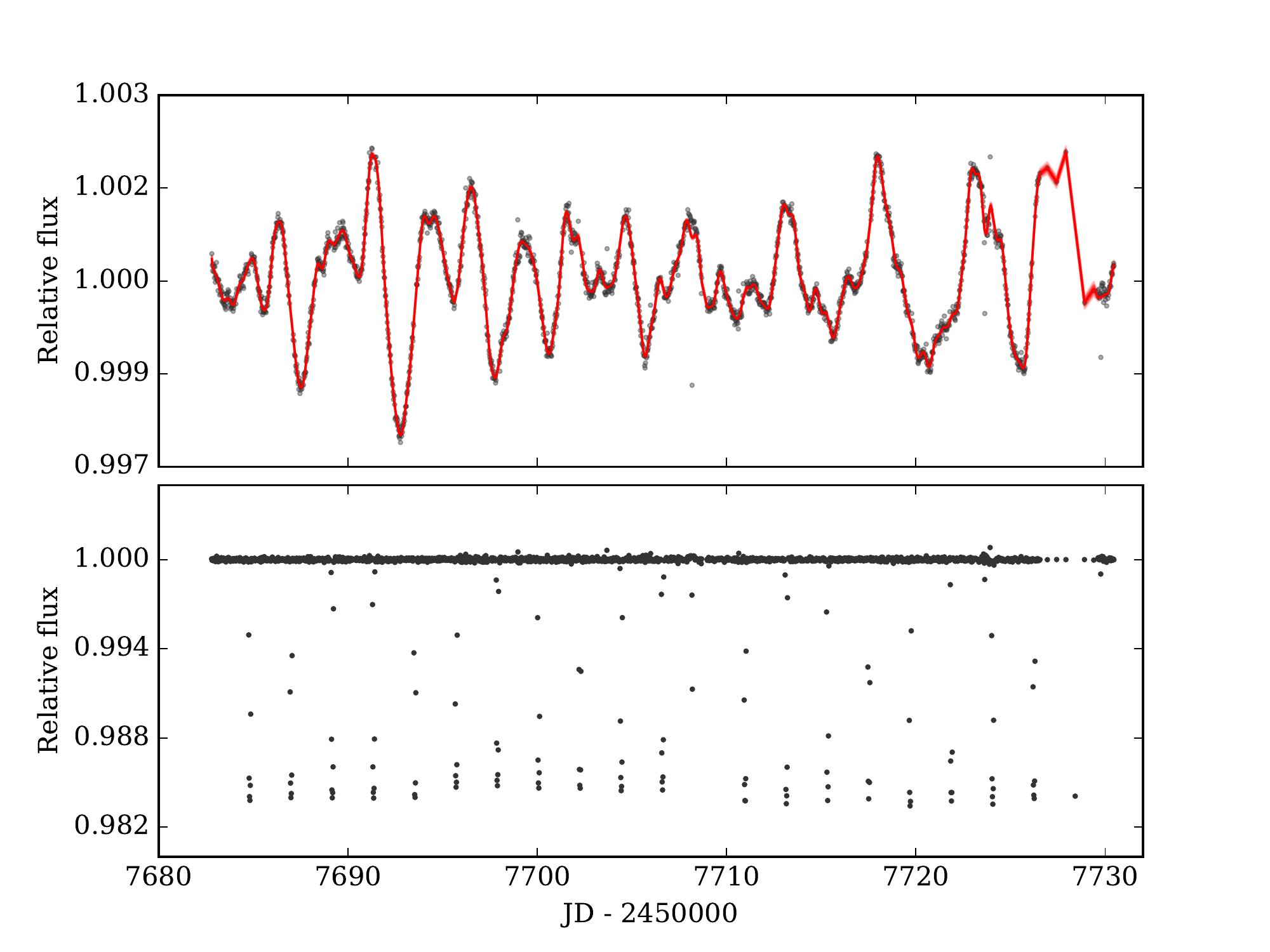}
\caption{Top panel: Detrended light curve, with the transits masked out (black points). The red line represents the GP adjusted to the data, using the most probable hyper-parameters from the MCMC. Bottom panel: Final light curve, with the most probable GP fit removed.}
\label{fig:GP_lc}
\end{figure}

It is possible to measure the rotational period of a star from its light curve. If one assumes that the star's surface contains spots blocking part of its flux, then a periodic signal will be produced and it can be detected in the light curve. This effect can be spotted in the data of EPIC229426032. The rotational period can be measured by using the autocorrelation function (ACF), which has been used with Kepler data in the literature \citep[e.g.][]{McQuillan2013, Lopez-morales2016, Giles2017}. For this analysis we used the final light curve obtained from Section \ref{sec:photometry}, after detrending and removing the long term variation.
 
We produced the ACF by following the method described in \citet{Edelson1988}, using the implementation from \texttt{astroML}\footnote{\url{http://www.astroml.org/modules/generated/astroML.time_series.ACF_EK.html}}. 
\citet{Lopez-morales2016} showed that the ACF follows a behavior similar to that of an under-damped simple harmonic oscillator (uSHO):

\begin{equation}\label{eq:uSHO}
y(t) = e^{-t\text{/}\tau_\mathrm{AR}}\left[A\cos{\left(\frac{2\pi t}{P}\right)} + B\cos{\left(\frac{4\pi t}{P}\right)}\right] + y_0,
\end{equation}

where $\tau_\mathrm{AR}$ is the decay timescale, $P$ is the rotation period, both in units of days, and $y_0$ is a constant. 

We fit equation \ref{eq:uSHO} to our ACF using a least-square minimization, and obtained the following solutions: $\tau_{AR}=9.0 \pm 0.6$~days, $P=5.07 \pm 0.02$~days, $A=0.59 \pm 0.02$, $B=0.32 \pm 0.02$, and $y_0=0.02 \pm 0.01$. Therefore, these results provide an rotational period of the star of $P=5.07 \pm 0.02$~days. The result is shown in Figure \ref{fig:ACF_odr}. 

Before further analysis of the light curve to search for transit signals, we had to remove the effect of rotational modulation from the data. This was done through Gaussian Process (GP) analysis. Several works \citep[e.g.][]{Vanderburg2015,Aigrain2016, Angus2018} have shown that a Quasi-Periodic kernel can model sinusoidal variations in a dataset, with decay components.
The Quasi-Periodic kernel is defined as: 

\begin{equation}\label{eq:GP}
k(t,t') = a^2 \exp \left[-\frac{(t-t')^2}{\theta_1^2} - \frac{1}{\theta_2^2}\sin^2\left(\frac{\pi(t-t')}{P}\right)\right],
\end{equation}

where $a$ is the amplitude of the covariance function, $\theta_1$ is the time scale of the exponential decay, $\theta_2$ and $P$ are the amplitude and period of the sinusoidal component. We also included a white noise component to the kernel, of the form $\sigma_n^2\, \delta_{t,t'}$, where $\delta_{t,t'}$ is the Kronecker delta. The values obtained from the ACF analysis were used as priors for $P$ and $\theta_1$ \citep{Haywood2014,Lopez-morales2016}. The amplitude was set to be constrained by the amplitude of the data, and $\theta_2$ to be within 0.05 and 5.0, following \citet{Jeffers2009}. The priors and best-fit values for each quantity are listed in Table~\ref{tab:kernel_parameters}.

\begin{table}
	\centering
	\caption{Priors and best-fit results obtained for the quasiperiodic kernel parameters.}
	\label{tab:kernel_parameters}
    \begin{threeparttable}
    {\renewcommand{\arraystretch}{1.7}
    \setlength{\tabcolsep}{15pt}
	\begin{tabular}{lcc}
		\hline
        \hline
        Parameter & Prior\tnote{a} & Best-fit value\tnote{b}\\
        \hline
        $a$ & $\mathcal{J}(0.0001, 0.5)$ & $0.249^{+0.017}_{-0.020}$\\
        $\theta_1$ & $\mathcal{N}(9.0, 0.6)$ & $9.105^{+0.064}_{-0.065}$\\
        $\theta_2$ & $\mathcal{N}(0.5, 0.05)$ & $0.575^{+0.010}_{-0.009}$\\
        $P$ & $\mathcal{N}(5.07, 0.02)$ & $5.11^{+0.02}_{-0.02}$\\
        $\ln \sigma_n^2$ & $\mathcal{J}(-15, -6)$ & $14.994^{+0.011}_{-0.004}$\\
        \hline
		\hline
    \end{tabular}}\quad
    \begin{tablenotes}
    	\item[a] {\footnotesize $\mathcal{N}(\mu, \sigma)$ represents a normal prior with with mean $\mu$ and standard deviation $\sigma$. $\mathcal{J}(a,b)$ represents a Jeffrey's prior with limits $a$ and $b$.}
        \item[b] {\footnotesize The values are shown as $B^{C-B}_{B-A}$, where $A$, $B$ and $C$ correspond to the 16, 50 and 84\% percentiles.}
      \end{tablenotes}
\end{threeparttable}
\end{table}

We used the \texttt{george}\footnote{\url{http://dan.iel.fm/george/current/}} implementation of GP analysis, along with the \texttt{emcee} package, to adjust this kernel to our data by performing an MCMC sampling. The posterior distributions for each parameter of the Quasi-periodic kernel are shown in Figure \ref{fig:CL002_GP}. 
The final fit to the light curve is shown in Figure \ref{fig:GP_lc}. The resulting light curve, without the effect of stellar rotation, was then used to derive the planet parameters for this star. 
Using the rotational period, with the stellar radius and the projected rotational velocity from Table~\ref{tab:properties}, we obtain the rotational velocity and star inclination to be $v_{\text{rot}} = 14.31^{+0.59}_{-0.67}\,\kmpers$ and $i = 51.56^{+3.73}_{-2.80}$~degrees.
For EPIC246067459, we could not measure the rotational period using this method because the signal by the stellar rotation embedded in the light curve was not as strong as with the other star.

\subsection{Joint Analysis}\label{sec:joint_analysis}

\begin{table*}
	\centering
	\caption{Physical and orbital parameters for both planets, derived from the results from the \texttt{exonailer} run.}
	\label{tab:exonailer_output}
    \begin{threeparttable}
    {\renewcommand{\arraystretch}{1.7}
    \setlength{\tabcolsep}{12pt}
	\begin{tabular}{llcccc}
		\hline
        \hline
        & & \multicolumn{2}{c}{EPIC229426032~$b$} & \multicolumn{2}{c}{EPIC246067459~$b$}\\
		Parameter & Unit & Prior\tnote{a} & Best-fit value\tnote{b} & Prior\tnote{a} & Best-fit value\tnote{b}\\
        \hline
        Period & days & $\mathcal{N}(2.18057,0.1)$ & $2.18056^{+0.00002}_{-0.00002}$ & $\mathcal{N}(3.20466,0.1)$ & $3.20466^{+0.00003}_{-0.00003}$ \\
        $T_0$ - 2450000 &days & $\mathcal{N}(7684.8101,0.1)$ & $7684.8101^{+0.0001}_{-0.0001}$ & $\mathcal{N}(7740.5036,0.1)$ & $7740.5036^{+0.0004}_{-0.0004}$ \\
        $a/$\Rstar & & $\mathcal{U}(1,300)$ & $5.50^{+0.15}_{-0.11}$ & $\mathcal{U}(1,300)$ & $6.27^{+0.66}_{-0.52}$ \\
        $R_{\rm P}/$\Rstar & & $\mathcal{U}(0.001,0.5)$ & $0.118^{+0.001}_{-0.002}$ & $\mathcal{U}(0.001,0.5)$ & $0.080^{+0.003}_{-0.002}$ \\
        $i$  & deg & $\mathcal{U}(0,90)$ & $84.3^{+0.7}_{-0.4}$ & $\mathcal{U}(0,90)$ & $84.5^{+1.8}_{-1.5}$ \\
        $q_{1}$\tnote{c} & & $\mathcal{U}(0,1)$ & $0.15^{+0.10}_{-0.04}$ & $\mathcal{U}(0,1)$ & $0.53^{+0.29}_{-0.22}$ \\
        $q_{2}$\tnote{c} & & $\mathcal{U}(0,1)$ & $0.69^{+0.21}_{-0.26}$ & $\mathcal{U}(0,1)$ & $0.28^{+0.30}_{-0.15}$ \\ 
        $\sigma_w$ & ppm & $\mathcal{J}(10,500)$ & $128.2^{+2.8}_{-2.6}$ & $\mathcal{J}(10,500)$ & $369.9^{+4.6}_{-4.6}$ \\
        $K$ & \kmpers & $\mathcal{N}(0.3,0.1)$ & $0.21^{+0.01}_{-0.01}$ & $\mathcal{N}(0.1,0.1)$ & $0.10^{+0.01}_{-0.01}$ \\
        $e$ && fixed & $0.0$ & fixed & $0.0$ \\
        $\omega$ & deg & fixed & $90$ & fixed & $90$\\
        $\mu_{\rm CORALIE}$& \kmpers & $\mathcal{N}(-22.3,0.05)$ & $-22.27^{+0.03}_{-0.03}$ & & - \\      
        CORALIE jitter & \kmpers & $\mathcal{J}(0.0001,1)$ & $0.09^{+0.03}_{-0.02}$ & & -\\ 
        $\mu_{\rm HARPS}$& \kmpers & $\mathcal{N}(-22.3,0.05)$ & $-22.26^{+0.01}_{-0.02}$ & &  -\\      
        HARPS jitter & \kmpers & $\mathcal{J}(0.0001,1)$ & $0.002^{+0.015}_{-0.002}$ & & - \\
        $\mu_{\rm FEROS}$& \kmpers & & - & $\mathcal{N}(8.22,0.05)$ & $8.26^{+0.01}_{-0.01}$ \\      
        FEROS jitter & \kmpers & & - & $\mathcal{J}(0.0001,0.1)$ & $0.03^{+0.01}_{-0.01}$ \\
        
        \hline
        $M_{\rm P}$ & \mjup & & $1.60^{+0.11}_{-0.11}$ & & $0.86^{+0.13}_{-0.12}$\\
		$R_{\rm P}$\tnote{d} & \rjup & & $1.65^{+0.07}_{-0.08}$ & & $1.30^{+0.15}_{-0.14}$ \\
		$\rho_{\rm P}$ & g cm$^{-3}$ & & $0.44^{+0.08}_{-0.06}$ & & $0.56^{+0.25}_{-0.16}$\\
		$a$ & AU & & $0.037^{+0.002}_{-0.002}$ & & $0.046^{+0.007}_{-0.006}$\\
		$T_{\rm eq.}$ & K & & $1884^{+37}_{-36}$ & & $1587^{+75}_{-76}$\\
		<$F$>\tnote{e} & $10^9\,\fluxcgs$ & & $2.86^{+0.23}_{-0.21}$ & & $1.44^{+0.29}_{-0.26}$\\
		$H$\tnote{f} & $10^8$ cm & & $1.06^{+0.13}_{-0.12}$ & & $0.93^{+0.21}_{-0.22}$\\
        
        \hline
		\hline
    \end{tabular}}\quad
    \begin{tablenotes}
    	\item[a] {\footnotesize $\mathcal{N}(\mu, \sigma)$ represents a normal prior with with mean $\mu$ and standard deviation $\sigma$. $\mathcal{U}(a,b)$ represents an uniform prior with limits $a$ and $b$. $\mathcal{J}(a,b)$ represents a Jeffrey's prior with limits $a$ and $b$.}
        \item[b] {\footnotesize The values are shown as $B^{C-B}_{B-A}$, where $A$, $B$ and $C$ correspond to the 16, 50 and 84\% percentiles.}
        \item[c] {\footnotesize $q_1$ and $q_2$ are the sampling coefficients to fit for a quadratic limb-darkening law, defined in \citet{Kipping2013}. The limb-darkening coefficients can be recovered as $\mu_1 = 2\sqrt{q_1}q_2$ and $\mu_2 = \sqrt{q_1}(1-2q_2)$.}
        \item[d] {\footnotesize The planet radius for EPIC246067459~$b$ considers the transit depth and the dilution produced by nearby stars (section \ref{sec:AO_imaging}). The uncorrected radius was found to be $1.24^{+0.13}_{-0.14}\,\rjup$.}
        \item[e] {\footnotesize Orbit averaged incident flux.}
        \item[f] {\footnotesize Scale height, assuming hydrogen dominated composition.}
      \end{tablenotes}
\end{threeparttable}
\end{table*}

\begin{figure}
\centering
\includegraphics[width=8.5cm]{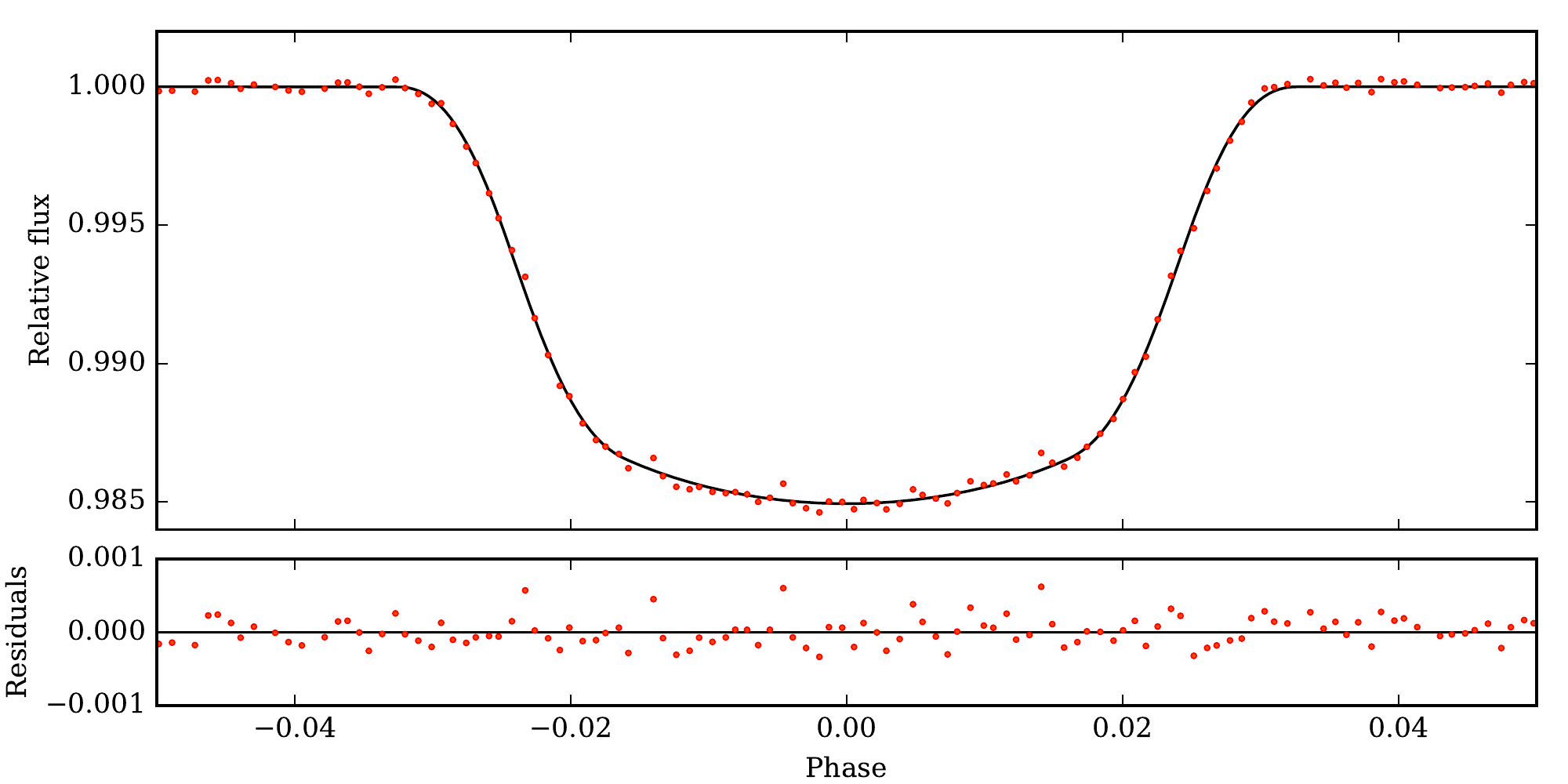}\\
\includegraphics[width=8.5cm]{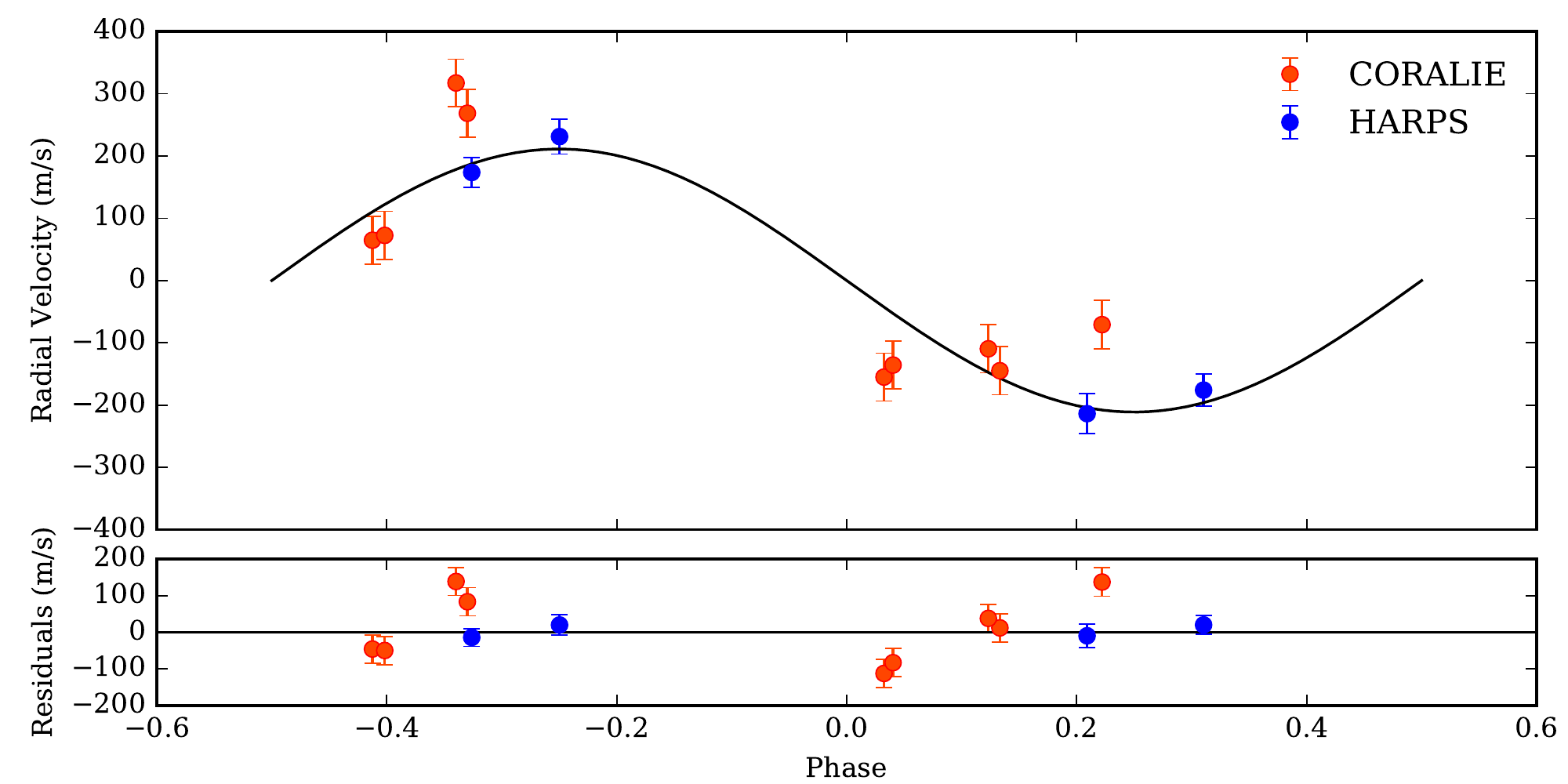}
\caption{\texttt{exonailer} fit for EPIC229426032. Top panel: Relative flux vs. orbital phase. Bottom panel: Radial Velocity data vs. phase, where red points represent Coralie, and blue points HARPS data. For both plots, the black lines represent the models with the most probable solution for the \texttt{exonailer} fit, with parameters listed in Table~\ref{tab:exonailer_output}.}
\label{fig:joint_fit_CL002}
\end{figure}

\begin{figure}
\centering
\includegraphics[width=8.5cm]{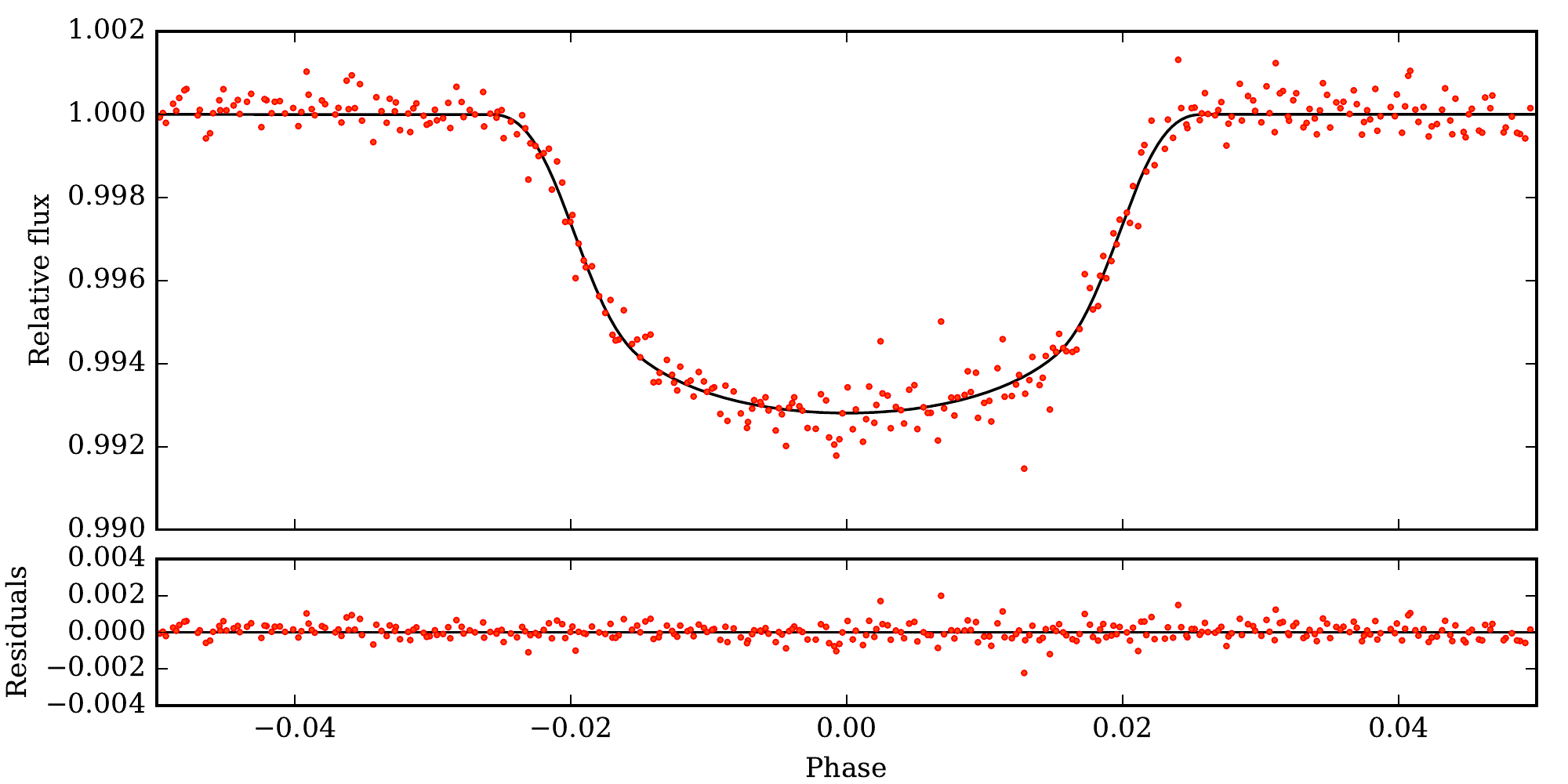}\\
\includegraphics[width=8.5cm]{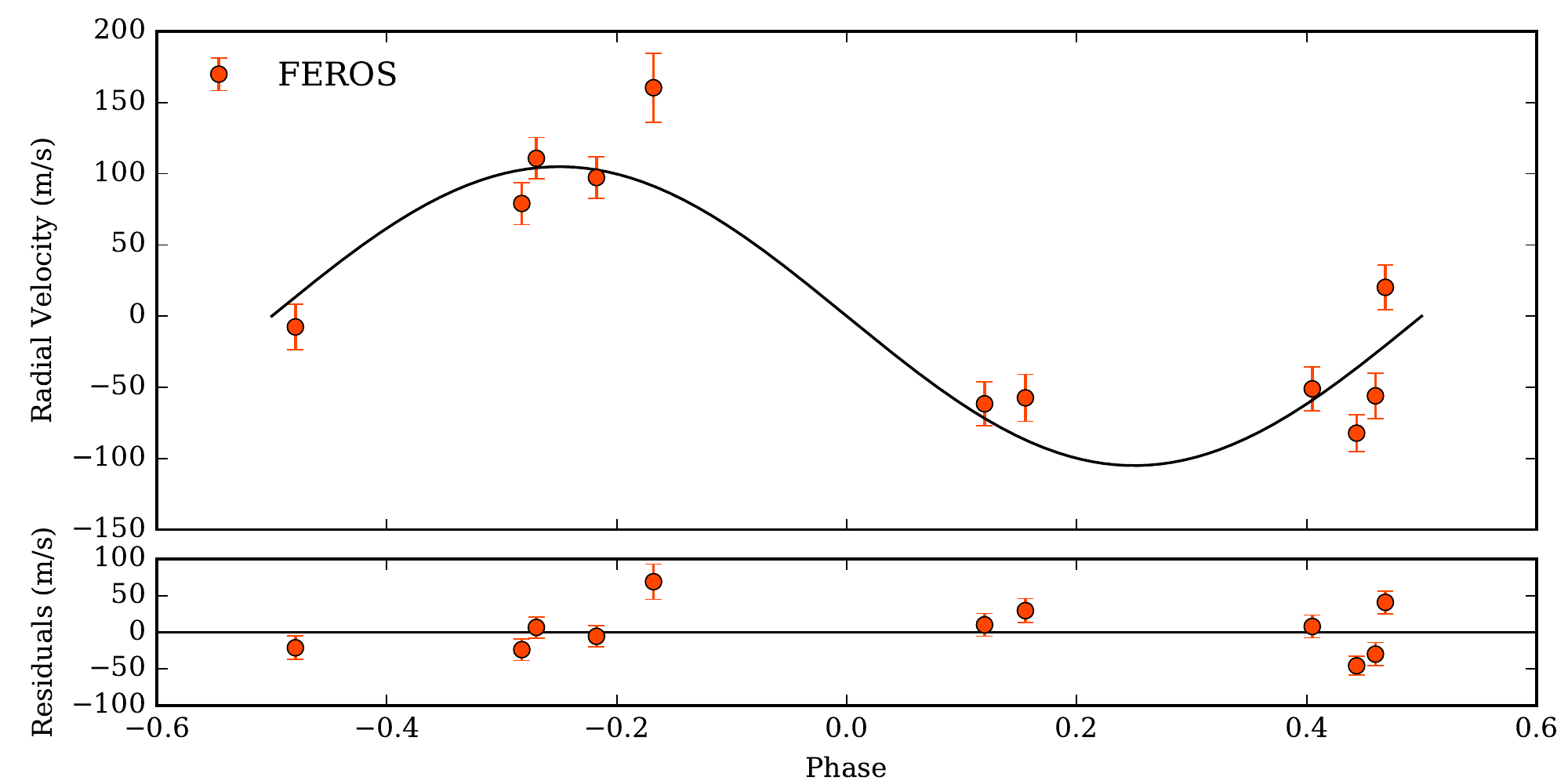}
\caption{\texttt{exonailer} fit for EPIC246067459. Top panel: Relative flux vs. orbital phase. Bottom panel: Radial Velocity data vs. phase, where red points represent FEROS data. For both plots, the black lines represent the models with the most probable solution for the \texttt{exonailer} fit, with parameters listed in Table~\ref{tab:exonailer_output}.}
\label{fig:joint_fit_CL003}
\end{figure}

In order to obtain a global solution for both systems, combining the photometry and radial velocity information, we used the \texttt{exonailer} code \citep{Espinoza2016}. \texttt{exonailer} is a tool that fits transit light curves, as well as radial velocity information, using a Bayesian approach to derive the most probable solution, for a given system, by using a set of priors for each one of the orbital and transit model parameters.
We used the quadratic limb-darkening law on both stars, which is the optimal one in our case following the algorithms and method detailed in \citet{Espinoza2016a}. We also fit for the limb-darkening coefficients instead of using modeled values, which has been shown to lead to important biases in the transit parameters \citep{Espinoza2015}.
We fitted the data of EPIC229426032 with both circular and non-circular models, and obtained that the eccentricity of the non-circular model was consistent with zero. The Bayesian Information Criterion (BIC) obtained for the circular orbit (BIC = -20.54) was also smaller compared with the non-circular one (BIC = -15.17), leading us to finally adopt a circular orbit for the system. The same analysis was done for EPIC246067459, where we also adopted a circular model.
The obtained distributions for each parameter, as well as the limb-darkening sampling coefficients, are listed in Table~\ref{tab:exonailer_output}.
For EPIC229426032, we used the light curve obtained in section~\ref{sec:rot_period}, and shown in the bottom panel of Figure~\ref{fig:GP_lc}, with the effect of stellar rotation and long term variations removed. For EPIC246067459, we used the detrended light curve obtained in section~\ref{sec:photometry}, and shown in the bottom panel of Figure~\ref{fig:CL003_lc}.
The transit and radial velocity solutions, given the posterior values from Table \ref{tab:exonailer_output}, are shown in Figure \ref{fig:joint_fit_CL002} and \ref{fig:joint_fit_CL003} for EPIC229426032 and EPIC246067459, respectively.

Using the stellar mass and radius computed in Section \ref{sec:stellar_params}, along with the values from Table~\ref{tab:exonailer_output}, we estimate the planet mass and radius to be $1.60^{+0.11}_{-0.11}\,\mjup$ and $1.65^{+0.07}_{-0.08}\,\rjup$, respectively, for EPIC229426032~$b$. 
For EPIC246067459~$b$, we also had to consider the dilution in the transit depth produced by the two detected nearby companions. After correcting by this factor, we found the planet mass and radius to be $0.86^{+0.13}_{-0.12}\, \mjup$ and $1.30^{+0.15}_{-0.14}\,\rjup$, respectively.
These quantities, along with other parameters, are summarized in Table~\ref{tab:exonailer_output}.

\subsection{Activity indicators}

\begin{figure*}
\centering
\includegraphics[scale=0.56]{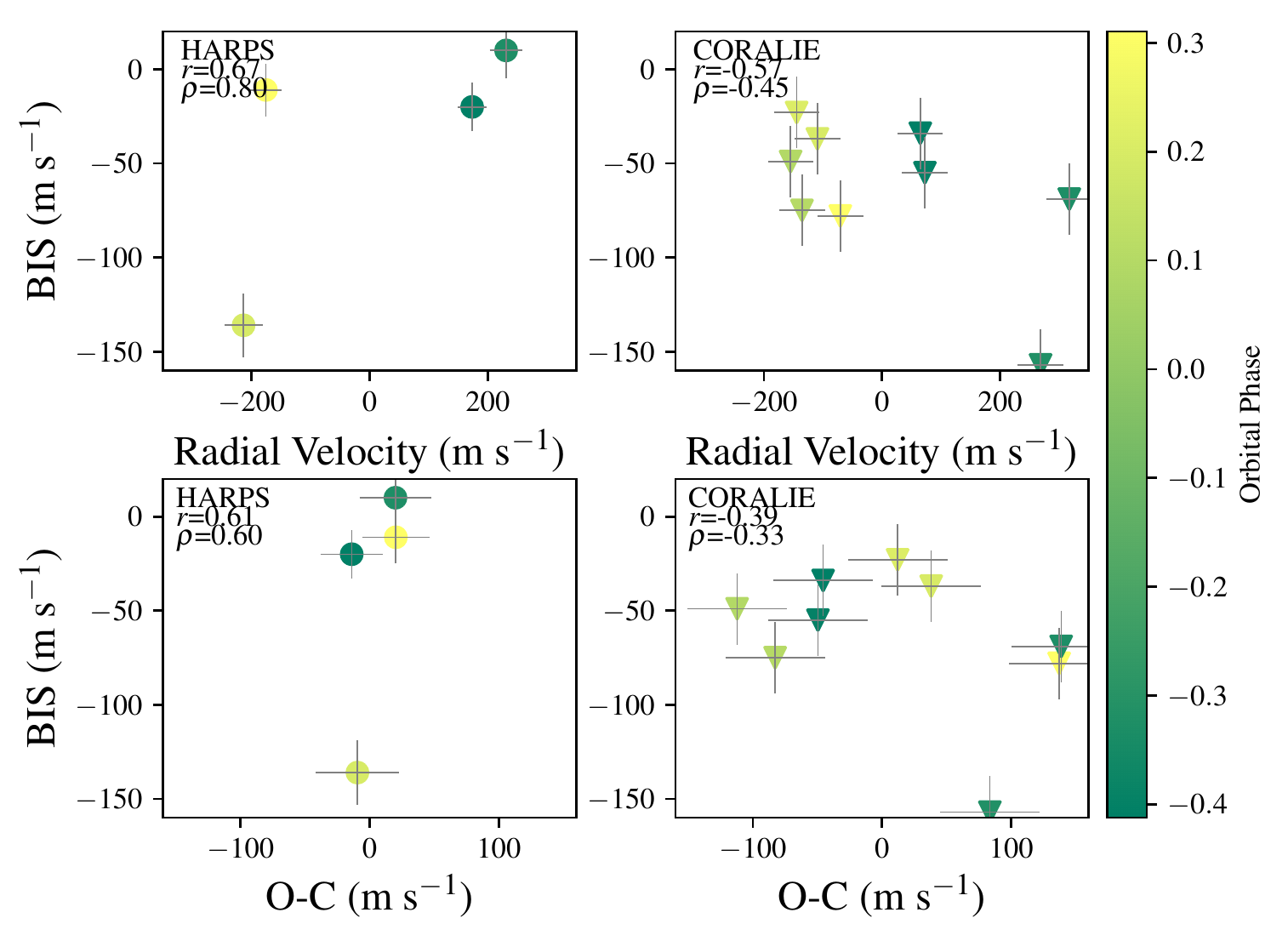}
\includegraphics[scale=0.56]{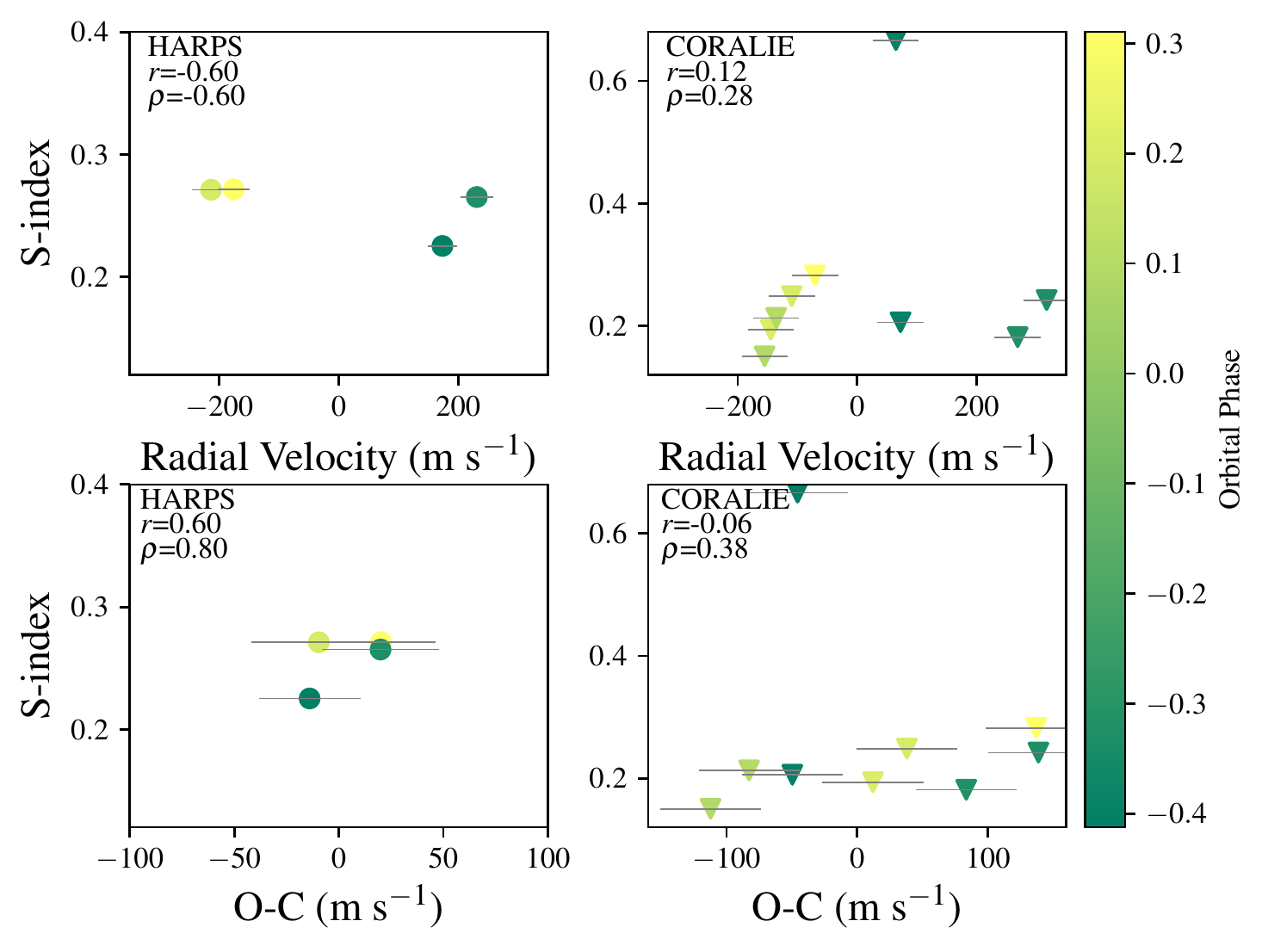}
\caption{Correlations between the RVs (top panels) and the activity indices available from both HARPS (left panels) and CORALIE (right panels) for EPIC229426032.  The lower panels show the same correlations for the residuals from the fit.  The colors represent the orbital phase of the RV and residuals. $r$ and $\rho$ are the Pearson and Spearman correlation coefficients, respectively.}
\label{fig:corr_cl002}
\end{figure*}

\begin{figure}
\centering
\includegraphics[scale=0.55]{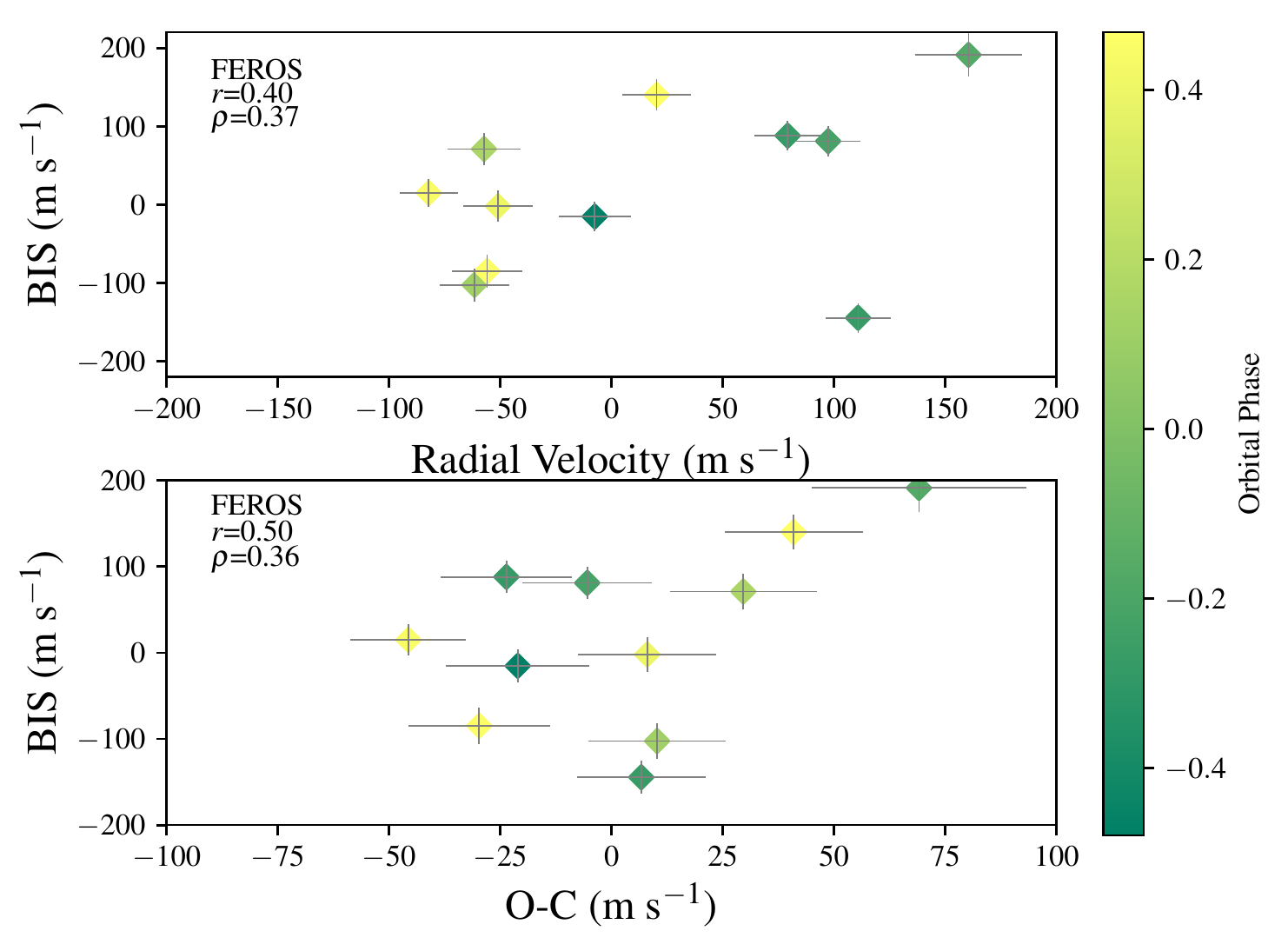}
\caption{Correlations between the radial velocities (top panel) and the residuals from the planetary fit (bottom panel), with the BIS from FEROS for EPIC246067459. The colors represent the orbital phase of the RV and residuals. $r$ and $\rho$ are the Pearson and Spearman correlation coefficients, respectively.}
\label{fig:corr_cl003}
\end{figure}

We measured a set of stellar activity indicators for both stars, in order to further confirm the planetary nature of the transit and radial velocity signals.
For EPIC229426032, we measured the Bisector Inverse Span \citep[BIS,][]{queloz2001,toner1988}, and the \ion{Ca}{ii} H and K S-index \citep{jenkins2008,jenkins2011}.
We used two coefficients to determine the level of correlation between the activity indices and the radial velocities for each instrument, the Pearson ($r$) and Spearman ($\rho$) correlation coefficients. 
For both quantities, the standard limits set for weak, moderate, and strong correlation between two quantities are $|r_c| < 0.5$, $0.5 \leq |r_c| \leq 0.7$, and $0.7 < |r_c|$, respectively.

For the HARPS data, we obtain $(r, \rho)_{\textrm{\footnotesize{BIS}}} = (0.67, 0.80)$, and $(r, \rho)_{\textrm{\footnotesize{S-index}}} = (0.60, 0.60)$, for the correlation between the BIS and the S-index with the RVs, respectively (Figure~\ref{fig:corr_cl002}). These results would suggest that both coefficients are correlated with the RVs, but the number of points considered is too small to make any robust conclusions. 
We performed the same analysis with the residuals from the planetary fit (see Figure~\ref{fig:joint_fit_CL002}), and obtained $(r, \rho)_{\textrm{\footnotesize{BIS}}} = (0.61, 0.60)$, and $(r, \rho)_{\textrm{\footnotesize{S-index}}} = (0.60, 0.80)$. This would also hint again at correlation with the activity indices, but as before, the number of points is too low to conclude whether this means there is moderate correlation between the quantities or not.

For the Coralie data, we find $(r, \rho)_{\textrm{\footnotesize{BIS}}} = (-0.57, -0.45)$, and $(r, \rho)_{\textrm{\footnotesize{S-index}}} = (0.12, 0.28)$. For the BIS, the coefficients would suggest weak to moderate correlation with the RVs. We find that this correlation is powered only by one point (RV = 270 m~s$^{-1}$, BIS = -157 m~s$^{-1}$), and if we remove it, the correlation drops to $(r, \rho)_{\textrm{\footnotesize{BIS}}} = (-0.24, -0.29)$.  This reality is confirmed by a jacknife-like analysis that moved through the data, removing individual points and re-performing the correlation tests, highlighting that only when this outlying data point is removed does the correlation coefficient change.  Too much statistical weight is being given to this one outlier.  In fact, when we combine the HARPS and Coralie measurements, the coefficients also drop into the weakly correlated category, showing that stellar activity may be impacting the RVs, but only by adding random noise.

In the case of the correlation with the residuals from the planet fit we obtain $(r, \rho)_{\textrm{\footnotesize{BIS}}} = (-0.39, -0.33)$, and $(r, \rho)_{\textrm{\footnotesize{S-index}}} = (-0.06, 0.38)$, which indicates no correlation among these quantities. These results, for the HARPS and Coralie data, can be seen in Figure~\ref{fig:corr_cl002}, with the activity indices listed in Table~\ref{tab:rvs_coralie} and \ref{tab:rvs_harps}. 

We also performed the bisector analysis on EPIC246067459, and found $(r, \rho)_{\textrm{\footnotesize{BIS}}} = (0.40, 0.37)$, which would indicate no correlation between the BIS and the FEROS RVs. For the residuals we found $(r, \rho)_{\textrm{\footnotesize{BIS}}} = (0.50, 0.36)$, also indicative of no strong correlation. The results are shown in Figure~\ref{fig:corr_cl003}, and listed in Table~\ref{tab:rvs_feros}. We did not include the S-index due to the low S/N spectra obtained with FEROS, which prohibited us from measuring them reliably.

\subsection{Planet scenario validation}

In order to confirm the planetary nature of our photometric and spectroscopic measurements, 
we performed a blend analysis using the algorithms described in \cite{hartman:2011,hartman:2011a}, 
which model the observations taking into account the possibility that they could be generated 
by either a planet, stellar companions physically associated with our target star or by 
various blend scenarios, including blended eclipsing binary and hierarchical triple systems.

EPIC 229426032b is confirmed to be a planet based solely on the photometry; it is practically 
impossible for the best-fit blend scenarios to fit the observed photometry in 
any of the cases consistent with the spectroscopic information. For EPIC 
246067459b, the planetary interpretation is also favored by the data: although there is a 
detected close-by companion in the Lick 3m AO data, the lightcurve is not consistent with the 
transit/eclipses arising from the neighbor, as all the simulated lightcurve signatures imply 
$J-K$ colors much less than the observed $J-K = 0.631 \pm 0.043$. Considering that the brighter 
source could still itself be a blend, we can reject all the blend scenarios at 2.5 
sigma-confidence based on the photometry. However, none of them are able to produce the observed 
$100$ m/s sinusoidal RV variation. The best-fit blend scenarios to the photometry also yield 
large bisector span variations in excess of 1 km/s, which are clearly ruled out by our 
measurements (see Figure \ref{fig:corr_cl003}). We consider thus both planets to be statistically 
validated given our photometric and spectroscopic measurements.

\subsection{Searching for additional signals in the photometry}

We search for additional signals in our K2 light curves, produced by other companions, orbital phase variations, or secondary eclipses by performing a Box-fitting Least Squares periodogram \citep[BLS,][]{BLS} on the light curves, with the transits of the detected planets removed. We find no significant peak in the BLS for both stars, which limits the transit depth of the possible additional companions to be less than 220~ppm and 250~ppm for EPIC229426032 and EPIC246067459, respectively, for a $3\sigma$ detection.
We could not detect secondary eclipses in neither of the light curves. For EPIC229426032 we had placed an upper limit for the depth of the eclipse to be $(\Rp/a)^2 < 478$ ppm, so the fact that we could not detect it points to a geometric albedo of $A_\text{g} < 0.46$. This is in agreement with what has been found for hot Jupiters \citep{Heng2013,Esteves2015}. For EPIC246067459, it comes to no surprise that we could not detect its eclipse, given that its depth would have been $(\Rp/a)^2 < 163$~ppm, which is bellow the detection limit of the data. We could not detect orbital phase variations in neither of the light curves.

\section{Discussion}\label{sec:discussion}

\begin{figure}
\includegraphics[width=9cm]{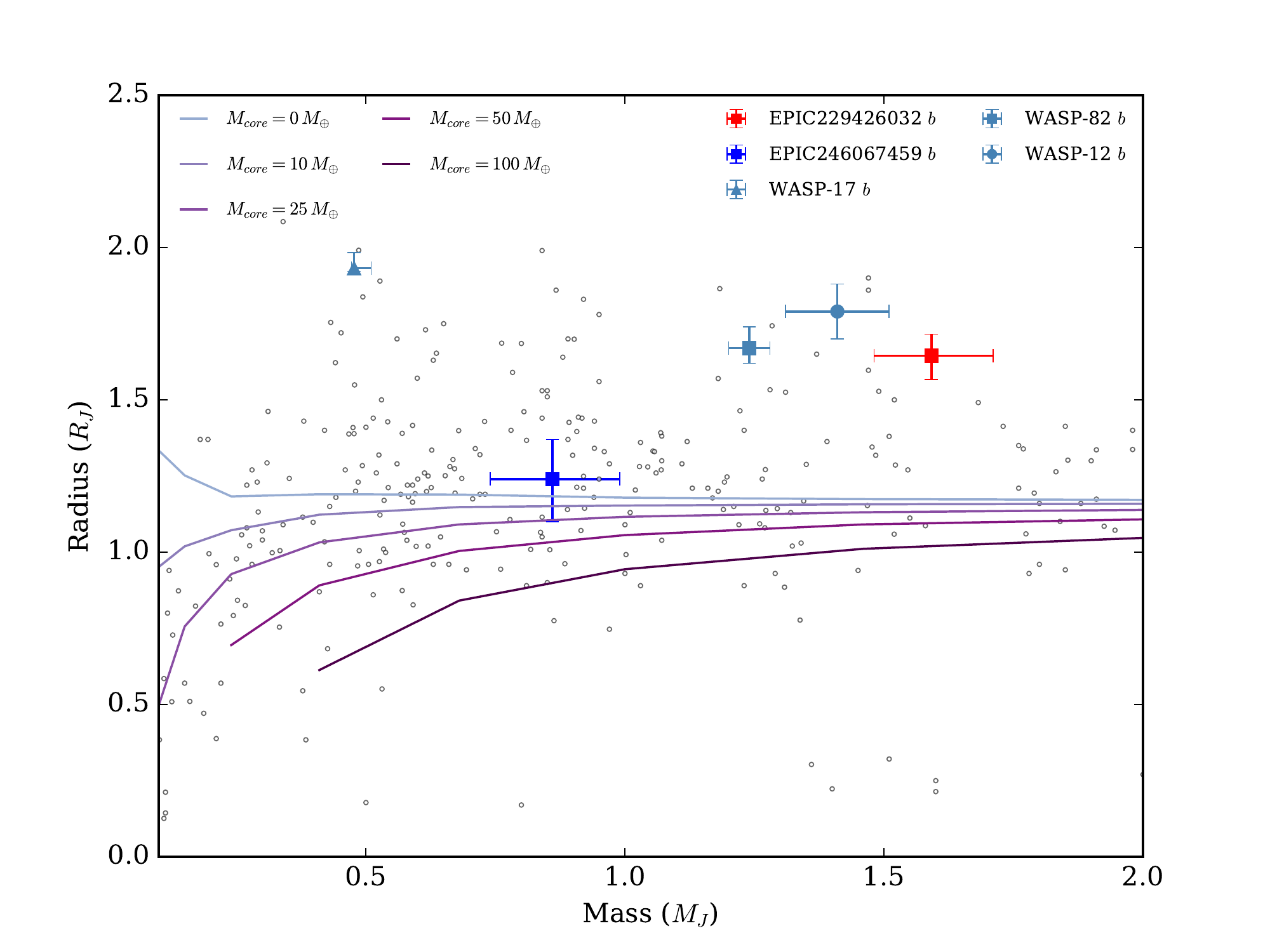}
\caption{Mass-radius diagram. The red and blue squares represent the planets detected in this work. The lines represent models from \citet{Fortney2007}, for hydrogen-helium dominated planets, at 0.02 AU from the parent star, and different values of core mass.
Light blue symbols represent inflated and very dense hot Jupiters mentioned in the text.
The white circles represent planets from the NASA Exoplanet archive,
with known mass and radius values, orbital period less than 10 days, and masses within 0.1-2.0 \mjup.}
\label{fig:mass_radius_plot}
\end{figure}

\begin{figure}
\includegraphics[width=9cm]{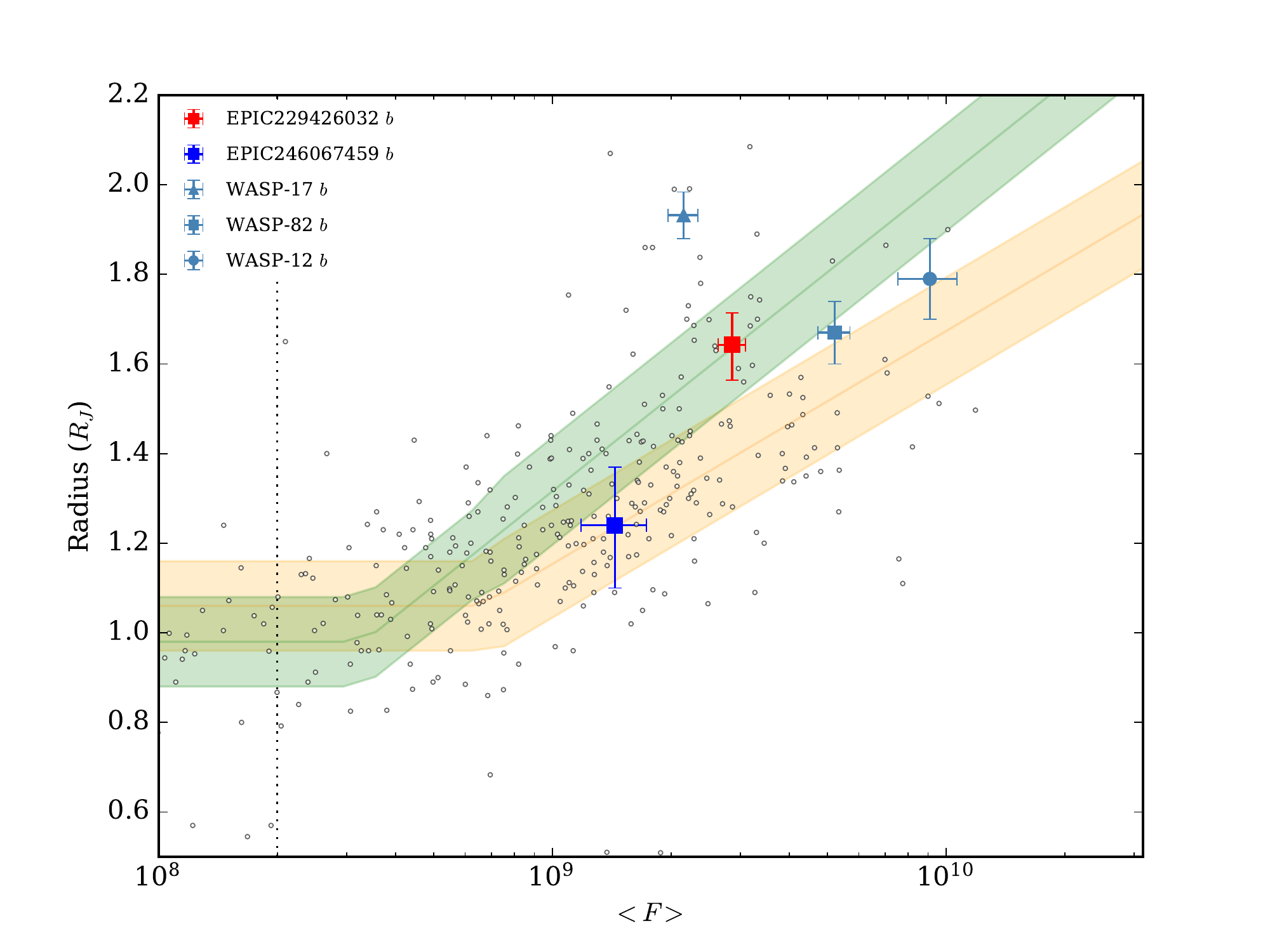}
\caption{Radius vs. planet incident flux. The red and blue squares represent the planets detected in this work. The light blue markers and white circles represent the same systems as in Figure~\ref{fig:mass_radius_plot}. The vertical dotted line represents the flux threshold $2 \times 10^8\,\fluxcgs$, above which planets have been found to be inflated. The shaded areas represent the relations from \citet{Sestovic2018} for $M = 0.37-0.98\,\mjup$ (green) and $M = 0.98-2.50\,\mjup$ (orange), and their $1\sigma$ confidence levels.}
\label{fig:radius_flux}
\end{figure}

We compared the mass and radius of both planets with the models from \citet{Fortney2007}, for hydrogen-helium dominated planets, with different amounts of metal compositions (represented by the core mass). We found for EPIC229426032~$b$ that the radius is significantly higher than expected for the given mass (0.5\,\rjup\, larger than the model for a 4.5~Gyr-old planet with semi-major axis of 0.02~AU and no core). This is shown in Figure \ref{fig:mass_radius_plot}. 
We looked at the confirmed planets list from the NASA Exoplanet Archive\footnote{\url{https://exoplanetarchive.ipac.caltech.edu/index.html}}, and found that EPIC 229426032~$b$ falls into a region of highly inflated hot Jupiters that is as yet not very well populated. We also compared the planet with other cases of highly inflated hot Jupiters, like WASP-17~$b$ \citep{anderson2010}, WASP-82~$b$ \citep{west2016}, and WASP-12~$b$ \citep{hebb2009}. These planets have shown to be good cases to perform atmospheric studies, which makes EPIC 229426032~$b$ a good laboratory for studying the atmospheres of highly inflated planets as well.
For EPIC 246067459~$b$, we find its radius to be consistent with the models of \cite{Fortney2007} for a hydrogen and helium dominated planet with a core mass up to 25 $M_{\oplus}$, at the $1\sigma$ level.

As was mentioned in the introduction, some studies trying to detect the source of planetary inflation point at correlations between the planet's incident flux and radius \citep[e.g.][]{Demory2011, Laughlin2011}, and have detected an incident flux threshold $F_i = 2 \times 10^8\,\fluxcgs$, above which inflation is found to happen. Both of our planets fall above this threshold as shown in Figure~\ref{fig:radius_flux}, which suggests inflation is shaping the observed radius of our newly discovered exoplanets. 
We see that EPIC 229426032~b is considerably larger than what theoretical models predict for a H/He dominated planet, receiving high radiation levels. 
In the case of EPIC 246067459~$b$, its mass and radius seem to be consistent with it not-being inflated, even though it receives a high incident flux (Fig.~\ref{fig:radius_flux}). 
We also compared the two planets from this work with the models of radius against incident flux and mass by \citet{Sestovic2018}. 
Here we see again see that EPIC 229426032~b appears to be even more inflated than what the model from \citet{Sestovic2018} predicts ($\Mp = 0.98\text{--}2.50\,\mjup$, orange area in Figure~\ref{fig:radius_flux}). 
We also find that the scale height estimated for this planet (see Table~\ref{tab:exonailer_output}) is comparable to those of systems currently targeted for atmospheric characterization \citep[e.g., WASP-12b, with $H \sim 1100$~km, ][]{Burton2015}. The latter, and given that the planet orbits a bright host star, again makes EPIC 229426032~b appear to be an excellent candidate for follow-up studies.
For EPIC 246067459~$b$, we see that its radius is consistent with a non-inflated planet of mass within $0.37-0.98\, \mjup$ within $1\sigma$ (represented by the flat part of the green region in Figure~\ref{fig:radius_flux}).

\citet{jenkins2017} show that gas giant planets with orbital periods less than 100 days orbit stars that are significantly more metal-rich than their counterparts that host longer period giant planets. Furthermore, they also discovered a difference in the host star metallicity of Jupiter-mass planets and super-Jupiters, whereby the Jupiter-mass planets orbit stars significantly more metal-rich than those with significantly higher masses. This result was later confirmed at higher statistical significance by \citet{santos2017}. The very short period systems detected in this work also seem to orbit very metal-rich stars, and although the less massive of the two, EPIC226067459~$b$, is still classed as a Jupiter-mass gas giant for the purposes of the metallicity-mass relationship discovered by \citeauthor{jenkins2017}, it is intriguing that it orbits a significantly more metal-rich star than EPIC229426032~$b$.


\section{Summary}\label{sec:summary}

We present the discovery of two new hot Jupiters from our Chilean K2 project that aims to detect new planets in the southern fields of the K2 mission. For EPIC229426032~$b$, our best solution is consistent with a hot Jupiter planet with a $R = 1.65\,\rjup$, orbiting its host star in a period of 2.2 days. Its radius makes it a highly inflated hot Jupiter, and when coupled with the brightness of the host, it makes an excellent candidate for further atmospheric studies.

EPIC246067459~$b$, on the other hand, appears to have a mass similar to that of Jupiter, a radius of $R = 1.30\,\rjup$, and orbital period of 3.2 days. Even though this planet is in the regime where planetary inflation is important, it was found to have a radius consistent with theoretical models for H/He dominated objects.

\section*{Acknowledgements}

MGS acknowledges the support of CONICYT-PFCHA/Doctorado Nacional-21141037, Chile. MRD is supported by CONICYT-PFCHA/Doctorado Nacional-21140646, Chile. JSJ acknowledges support by FONDECYT grant 1161218 and partial support by CATA-Basal (PB06, CONICYT). A.J.\ acknowledges support from FONDECYT project 1171208, BASAL CATA PFB-06, and by the Ministry for the Economy, Development, and Tourism's Programa Iniciativa Cient\'{i}fica Milenio through grant IC\,120009, awarded to the Millennium Institute of Astrophysics (MAS). 
H.D.\ acknowledges support from FONDECYT Postdoctorado 3150314 and FONDECYT Regular 1171364, from Fondo Nacional de Desarrollo Cient\'ifico y Tecnol\'ogico.
R.L. acknowledges support from BASAL CATA PFB-06.




\bibliographystyle{mnras}
\bibliography{refs_paper} 

\begin{thebibliography}{}
\makeatletter
\relax
\def\mn@urlcharsother{\let\do\@makeother \do\$\do\&\do\#\do\^\do\_\do\%\do\~}
\def\mn@doi{\begingroup\mn@urlcharsother \@ifnextchar [ {\mn@doi@}
  {\mn@doi@[]}}
\def\mn@doi@[#1]#2{\def\@tempa{#1}\ifx\@tempa\@empty \href
  {http://dx.doi.org/#2} {doi:#2}\else \href {http://dx.doi.org/#2} {#1}\fi
  \endgroup}
\def\mn@eprint#1#2{\mn@eprint@#1:#2::\@nil}
\def\mn@eprint@arXiv#1{\href {http://arxiv.org/abs/#1} {{\tt arXiv:#1}}}
\def\mn@eprint@dblp#1{\href {http://dblp.uni-trier.de/rec/bibtex/#1.xml}
  {dblp:#1}}
\def\mn@eprint@#1:#2:#3:#4\@nil{\def\@tempa {#1}\def\@tempb {#2}\def\@tempc
  {#3}\ifx \@tempc \@empty \let \@tempc \@tempb \let \@tempb \@tempa \fi \ifx
  \@tempb \@empty \def\@tempb {arXiv}\fi \@ifundefined
  {mn@eprint@\@tempb}{\@tempb:\@tempc}{\expandafter \expandafter \csname
  mn@eprint@\@tempb\endcsname \expandafter{\@tempc}}}

\bibitem[\protect\citeauthoryear{{Aigrain}, {Parviainen}  \& {Pope}}{{Aigrain}
  et~al.}{2016}]{Aigrain2016}
{Aigrain} S.,  {Parviainen} H.,   {Pope} B.~J.~S.,  2016, \mn@doi [\mnras]
  {10.1093/mnras/stw706}, \href
  {http://adsabs.harvard.edu/abs/2016MNRAS.459.2408A} {459, 2408}

\bibitem[\protect\citeauthoryear{{Anderson} et~al.,}{{Anderson}
  et~al.}{2010}]{anderson2010}
{Anderson} D.~R.,  et~al., 2010, \mn@doi [\apj] {10.1088/0004-637X/709/1/159},
  \href {http://adsabs.harvard.edu/abs/2010ApJ...709..159A} {709, 159}

\bibitem[\protect\citeauthoryear{{Angus}, {Morton}, {Aigrain}, {Foreman-Mackey}
   \& {Rajpaul}}{{Angus} et~al.}{2018}]{Angus2018}
{Angus} R.,  {Morton} T.,  {Aigrain} S.,  {Foreman-Mackey} D.,   {Rajpaul} V.,
  2018, \mn@doi [\mnras] {10.1093/mnras/stx2109}, \href
  {http://adsabs.harvard.edu/abs/2018MNRAS.474.2094A} {474, 2094}

\bibitem[\protect\citeauthoryear{{Borucki} et~al.,}{{Borucki}
  et~al.}{2010}]{Kepler}
{Borucki} W.~J.,  et~al., 2010, \mn@doi [Science] {10.1126/science.1185402},
  \href {http://adsabs.harvard.edu/abs/2010Sci...327..977B} {327, 977}

\bibitem[\protect\citeauthoryear{{Brahm} et~al.,}{{Brahm}
  et~al.}{2016}]{Brahm2016}
{Brahm} R.,  et~al., 2016, \mn@doi [\pasp] {10.1088/1538-3873/128/970/124402},
  \href {http://adsabs.harvard.edu/abs/2016PASP..128l4402B} {128, 124402}

\bibitem[\protect\citeauthoryear{{Brahm}, {Jord{\'a}n}  \& {Espinoza}}{{Brahm}
  et~al.}{2017a}]{Brahm2017}
{Brahm} R.,  {Jord{\'a}n} A.,   {Espinoza} N.,  2017a, \mn@doi [\pasp]
  {10.1088/1538-3873/aa5455}, \href
  {http://adsabs.harvard.edu/abs/2017PASP..129c4002B} {129, 034002}

\bibitem[\protect\citeauthoryear{{Brahm}, {Jord{\'a}n}, {Hartman}  \&
  {Bakos}}{{Brahm} et~al.}{2017b}]{Brahm2017b}
{Brahm} R.,  {Jord{\'a}n} A.,  {Hartman} J.,   {Bakos} G.,  2017b, \mn@doi
  [\mnras] {10.1093/mnras/stx144}, \href
  {http://adsabs.harvard.edu/abs/2017MNRAS.467..971B} {467, 971}

\bibitem[\protect\citeauthoryear{{Brahm} et~al.,}{{Brahm}
  et~al.}{2018}]{Brahm2018}
{Brahm} R.,  et~al., 2018, \mn@doi [\mnras] {10.1093/mnras/sty795}, \href
  {http://adsabs.harvard.edu/abs/2018MNRAS.tmp..775B} {}

\bibitem[\protect\citeauthoryear{{Burrows}, {Hubeny}, {Budaj}  \&
  {Hubbard}}{{Burrows} et~al.}{2007}]{burrows2007}
{Burrows} A.,  {Hubeny} I.,  {Budaj} J.,   {Hubbard} W.~B.,  2007, \mn@doi
  [\apj] {10.1086/514326}, \href
  {http://adsabs.harvard.edu/abs/2007ApJ...661..502B} {661, 502}

\bibitem[\protect\citeauthoryear{{Burton}, {Watson}, {Rodr{\'{\i}}guez-Gil},
  {Skillen}, {Littlefair}, {Dhillon}  \& {Pollacco}}{{Burton}
  et~al.}{2015}]{Burton2015}
{Burton} J.~R.,  {Watson} C.~A.,  {Rodr{\'{\i}}guez-Gil} P.,  {Skillen} I.,
  {Littlefair} S.~P.,  {Dhillon} S.,   {Pollacco} D.,  2015, \mn@doi [\mnras]
  {10.1093/mnras/stu2149}, \href
  {http://adsabs.harvard.edu/abs/2015MNRAS.446.1071B} {446, 1071}

\bibitem[\protect\citeauthoryear{{Casagrande}, {Ram{\'{\i}}rez},
  {Mel{\'e}ndez}, {Bessell}  \& {Asplund}}{{Casagrande}
  et~al.}{2010}]{Casagrande2010}
{Casagrande} L.,  {Ram{\'{\i}}rez} I.,  {Mel{\'e}ndez} J.,  {Bessell} M.,
  {Asplund} M.,  2010, \mn@doi [\aap] {10.1051/0004-6361/200913204}, \href
  {http://adsabs.harvard.edu/abs/2010A%26A...512A..54C} {512, A54}

\bibitem[\protect\citeauthoryear{{Castelli} \& {Kurucz}}{{Castelli} \&
  {Kurucz}}{2004}]{ATLAS9}
{Castelli} F.,  {Kurucz} R.~L.,  2004, ArXiv Astrophysics e-prints, \href
  {http://adsabs.harvard.edu/abs/2004astro.ph..5087C} {}

\bibitem[\protect\citeauthoryear{{Charbonneau}, {Brown}, {Latham}  \&
  {Mayor}}{{Charbonneau} et~al.}{2000}]{charbonneau2000}
{Charbonneau} D.,  {Brown} T.~M.,  {Latham} D.~W.,   {Mayor} M.,  2000, \mn@doi
  [\apjl] {10.1086/312457}, \href
  {http://adsabs.harvard.edu/abs/2000ApJ...529L..45C} {529, L45}

\bibitem[\protect\citeauthoryear{{Demarque}, {Woo}, {Kim}  \& {Yi}}{{Demarque}
  et~al.}{2004}]{Demarque2004}
{Demarque} P.,  {Woo} J.-H.,  {Kim} Y.-C.,   {Yi} S.~K.,  2004, \mn@doi [\apjs]
  {10.1086/424966}, \href {http://adsabs.harvard.edu/abs/2004ApJS..155..667D}
  {155, 667}

\bibitem[\protect\citeauthoryear{{Demory} \& {Seager}}{{Demory} \&
  {Seager}}{2011}]{Demory2011}
{Demory} B.-O.,  {Seager} S.,  2011, \mn@doi [\apjs]
  {10.1088/0067-0049/197/1/12}, \href
  {http://adsabs.harvard.edu/abs/2011ApJS..197...12D} {197, 12}

\bibitem[\protect\citeauthoryear{{Edelson} \& {Krolik}}{{Edelson} \&
  {Krolik}}{1988}]{Edelson1988}
{Edelson} R.~A.,  {Krolik} J.~H.,  1988, \mn@doi [\apj] {10.1086/166773}, \href
  {http://adsabs.harvard.edu/abs/1988ApJ...333..646E} {333, 646}

\bibitem[\protect\citeauthoryear{{Espinoza} \& {Jord{\'a}n}}{{Espinoza} \&
  {Jord{\'a}n}}{2015}]{Espinoza2015}
{Espinoza} N.,  {Jord{\'a}n} A.,  2015, \mn@doi [\mnras]
  {10.1093/mnras/stv744}, \href
  {http://adsabs.harvard.edu/abs/2015MNRAS.450.1879E} {450, 1879}

\bibitem[\protect\citeauthoryear{{Espinoza} \& {Jord{\'a}n}}{{Espinoza} \&
  {Jord{\'a}n}}{2016}]{Espinoza2016a}
{Espinoza} N.,  {Jord{\'a}n} A.,  2016, \mn@doi [\mnras]
  {10.1093/mnras/stw224}, \href
  {http://adsabs.harvard.edu/abs/2016MNRAS.457.3573E} {457, 3573}

\bibitem[\protect\citeauthoryear{{Espinoza} et~al.,}{{Espinoza}
  et~al.}{2016}]{Espinoza2016}
{Espinoza} N.,  et~al., 2016, \mn@doi [\apj] {10.3847/0004-637X/830/1/43},
  \href {http://adsabs.harvard.edu/abs/2016ApJ...830...43E} {830, 43}

\bibitem[\protect\citeauthoryear{{Esteves}, {De Mooij}  \&
  {Jayawardhana}}{{Esteves} et~al.}{2015}]{Esteves2015}
{Esteves} L.~J.,  {De Mooij} E.~J.~W.,   {Jayawardhana} R.,  2015, \mn@doi
  [\apj] {10.1088/0004-637X/804/2/150}, \href
  {http://adsabs.harvard.edu/abs/2015ApJ...804..150E} {804, 150}

\bibitem[\protect\citeauthoryear{{Fortney}, {Marley}  \& {Barnes}}{{Fortney}
  et~al.}{2007}]{Fortney2007}
{Fortney} J.~J.,  {Marley} M.~S.,   {Barnes} J.~W.,  2007, \mn@doi [\apj]
  {10.1086/512120}, \href {http://adsabs.harvard.edu/abs/2007ApJ...659.1661F}
  {659, 1661}

\bibitem[\protect\citeauthoryear{{Gavel} et~al.,}{{Gavel}
  et~al.}{2014}]{Gavel2014}
{Gavel} D.,  et~al., 2014, in Adaptive Optics Systems IV. p. 914805 (\mn@eprint
  {arXiv} {1407.8207}), \mn@doi{10.1117/12.2055256}

\bibitem[\protect\citeauthoryear{{Giles}, {Collier Cameron}  \&
  {Haywood}}{{Giles} et~al.}{2017}]{Giles2017}
{Giles} H.~A.~C.,  {Collier Cameron} A.,   {Haywood} R.~D.,  2017, \mn@doi
  [\mnras] {10.1093/mnras/stx1931}, \href
  {http://adsabs.harvard.edu/abs/2017MNRAS.472.1618G} {472, 1618}

\bibitem[\protect\citeauthoryear{{Giles} et~al.,}{{Giles}
  et~al.}{2018}]{Giles2018}
{Giles} H.~A.~C.,  et~al., 2018, \mn@doi [\mnras] {10.1093/mnras/stx3300},
  \href {http://adsabs.harvard.edu/abs/2018MNRAS.475.1809G} {475, 1809}

\bibitem[\protect\citeauthoryear{{Hartman} et~al.,}{{Hartman}
  et~al.}{2011a}]{hartman:2011a}
{Hartman} J.~D.,  et~al., 2011a, \mn@doi [\apj] {10.1088/0004-637X/728/2/138},
  \href {http://adsabs.harvard.edu/abs/2011ApJ...728..138H} {728, 138}

\bibitem[\protect\citeauthoryear{{Hartman} et~al.,}{{Hartman}
  et~al.}{2011b}]{hartman:2011}
{Hartman} J.~D.,  et~al., 2011b, \mn@doi [\apj] {10.1088/0004-637X/742/1/59},
  \href {http://adsabs.harvard.edu/abs/2011ApJ...742...59H} {742, 59}

\bibitem[\protect\citeauthoryear{{Haywood} et~al.,}{{Haywood}
  et~al.}{2014}]{Haywood2014}
{Haywood} R.~D.,  et~al., 2014, \mn@doi [\mnras] {10.1093/mnras/stu1320}, \href
  {http://adsabs.harvard.edu/abs/2014MNRAS.443.2517H} {443, 2517}

\bibitem[\protect\citeauthoryear{{Hebb} et~al.,}{{Hebb}
  et~al.}{2009}]{hebb2009}
{Hebb} L.,  et~al., 2009, \mn@doi [\apj] {10.1088/0004-637X/693/2/1920}, \href
  {http://adsabs.harvard.edu/abs/2009ApJ...693.1920H} {693, 1920}

\bibitem[\protect\citeauthoryear{{Heng} \& {Demory}}{{Heng} \&
  {Demory}}{2013}]{Heng2013}
{Heng} K.,  {Demory} B.-O.,  2013, \mn@doi [\apj]
  {10.1088/0004-637X/777/2/100}, \href
  {http://adsabs.harvard.edu/abs/2013ApJ...777..100H} {777, 100}

\bibitem[\protect\citeauthoryear{{Howell} et~al.,}{{Howell}
  et~al.}{2012}]{Howell2012}
{Howell} S.~B.,  et~al., 2012, \mn@doi [\apj] {10.1088/0004-637X/746/2/123},
  \href {http://adsabs.harvard.edu/abs/2012ApJ...746..123H} {746, 123}

\bibitem[\protect\citeauthoryear{{Howell} et~al.,}{{Howell} et~al.}{2014}]{K2}
{Howell} S.~B.,  et~al., 2014, \mn@doi [\pasp] {10.1086/676406}, \href
  {http://adsabs.harvard.edu/abs/2014PASP..126..398H} {126, 398}

\bibitem[\protect\citeauthoryear{{Jeffers} \& {Keller}}{{Jeffers} \&
  {Keller}}{2009}]{Jeffers2009}
{Jeffers} S.~V.,  {Keller} C.~U.,  2009, in {Stempels} E.,  ed.,  American
  Institute of Physics Conference Series Vol. 1094, 15th Cambridge Workshop on
  Cool Stars, Stellar Systems, and the Sun. pp 664--667,
  \mn@doi{10.1063/1.3099201}

\bibitem[\protect\citeauthoryear{{Jenkins}, {Jones}, {Pavlenko}, {Pinfield},
  {Barnes}  \& {Lyubchik}}{{Jenkins} et~al.}{2008}]{jenkins2008}
{Jenkins} J.~S.,  {Jones} H.~R.~A.,  {Pavlenko} Y.,  {Pinfield} D.~J.,
  {Barnes} J.~R.,   {Lyubchik} Y.,  2008, \mn@doi [\aap]
  {10.1051/0004-6361:20078611}, \href
  {http://adsabs.harvard.edu/abs/2008A%26A...485..571J} {485, 571}

\bibitem[\protect\citeauthoryear{{Jenkins} et~al.,}{{Jenkins}
  et~al.}{2011}]{jenkins2011}
{Jenkins} J.~S.,  et~al., 2011, \mn@doi [\aap] {10.1051/0004-6361/201016333},
  \href {http://adsabs.harvard.edu/abs/2011A%26A...531A...8J} {531, A8}

\bibitem[\protect\citeauthoryear{{Jenkins} et~al.,}{{Jenkins}
  et~al.}{2017}]{jenkins2017}
{Jenkins} J.~S.,  et~al., 2017, \mn@doi [\mnras] {10.1093/mnras/stw2811}, \href
  {http://adsabs.harvard.edu/abs/2017MNRAS.466..443J} {466, 443}

\bibitem[\protect\citeauthoryear{{Jones} et~al.,}{{Jones}
  et~al.}{2017}]{Jones2017}
{Jones} M.~I.,  et~al., 2017, preprint, \href
  {http://adsabs.harvard.edu/abs/2017arXiv170700779J} {} (\mn@eprint {arXiv}
  {1707.00779})

\bibitem[\protect\citeauthoryear{{Kaufer}, {Stahl}, {Tubbesing},
  {N{\o}rregaard}, {Avila}, {Francois}, {Pasquini}  \& {Pizzella}}{{Kaufer}
  et~al.}{1999}]{kaufer1999}
{Kaufer} A.,  {Stahl} O.,  {Tubbesing} S.,  {N{\o}rregaard} P.,  {Avila} G.,
  {Francois} P.,  {Pasquini} L.,   {Pizzella} A.,  1999, The Messenger, \href
  {http://adsabs.harvard.edu/abs/1999Msngr..95....8K} {95, 8}

\bibitem[\protect\citeauthoryear{{Kipping}}{{Kipping}}{2013}]{Kipping2013}
{Kipping} D.~M.,  2013, \mn@doi [\mnras] {10.1093/mnras/stt1435}, \href
  {http://adsabs.harvard.edu/abs/2013MNRAS.435.2152K} {435, 2152}

\bibitem[\protect\citeauthoryear{{Kov{\'a}cs}, {Zucker}  \&
  {Mazeh}}{{Kov{\'a}cs} et~al.}{2002}]{BLS}
{Kov{\'a}cs} G.,  {Zucker} S.,   {Mazeh} T.,  2002, \mn@doi [\aap]
  {10.1051/0004-6361:20020802}, \href
  {http://adsabs.harvard.edu/abs/2002A%26A...391..369K} {391, 369}

\bibitem[\protect\citeauthoryear{{Laughlin}, {Crismani}  \& {Adams}}{{Laughlin}
  et~al.}{2011}]{Laughlin2011}
{Laughlin} G.,  {Crismani} M.,   {Adams} F.~C.,  2011, \mn@doi [\apjl]
  {10.1088/2041-8205/729/1/L7}, \href
  {http://adsabs.harvard.edu/abs/2011ApJ...729L...7L} {729, L7}

\bibitem[\protect\citeauthoryear{{L{\'o}pez-Morales}
  et~al.,}{{L{\'o}pez-Morales} et~al.}{2016}]{Lopez-morales2016}
{L{\'o}pez-Morales} M.,  et~al., 2016, \mn@doi [\aj]
  {10.3847/0004-6256/152/6/204}, \href
  {http://adsabs.harvard.edu/abs/2016AJ....152..204L} {152, 204}

\bibitem[\protect\citeauthoryear{{Luger}, {Kruse}, {Foreman-Mackey}, {Agol}  \&
  {Saunders}}{{Luger} et~al.}{2017}]{Luger2017}
{Luger} R.,  {Kruse} E.,  {Foreman-Mackey} D.,  {Agol} E.,   {Saunders} N.,
  2017, preprint, \href {http://adsabs.harvard.edu/abs/2017arXiv170205488L} {}
  (\mn@eprint {arXiv} {1702.05488})

\bibitem[\protect\citeauthoryear{{Mayor} et~al.,}{{Mayor}
  et~al.}{2003}]{mayor2003}
{Mayor} M.,  et~al., 2003, The Messenger, \href
  {http://adsabs.harvard.edu/abs/2003Msngr.114...20M} {114, 20}

\bibitem[\protect\citeauthoryear{{McQuillan}, {Aigrain}  \&
  {Mazeh}}{{McQuillan} et~al.}{2013}]{McQuillan2013}
{McQuillan} A.,  {Aigrain} S.,   {Mazeh} T.,  2013, \mn@doi [\mnras]
  {10.1093/mnras/stt536}, \href
  {http://adsabs.harvard.edu/abs/2013MNRAS.432.1203M} {432, 1203}

\bibitem[\protect\citeauthoryear{{Miller} \& {Fortney}}{{Miller} \&
  {Fortney}}{2011}]{miller2011}
{Miller} N.,  {Fortney} J.~J.,  2011, \mn@doi [\apjl]
  {10.1088/2041-8205/736/2/L29}, \href
  {http://adsabs.harvard.edu/abs/2011ApJ...736L..29M} {736, L29}

\bibitem[\protect\citeauthoryear{{Queloz} et~al.,}{{Queloz}
  et~al.}{2000}]{Queloz2000}
{Queloz} D.,  et~al., 2000, \aap, \href
  {http://adsabs.harvard.edu/abs/2000A%26A...354...99Q} {354, 99}

\bibitem[\protect\citeauthoryear{{Queloz} et~al.,}{{Queloz}
  et~al.}{2001}]{queloz2001}
{Queloz} D.,  et~al., 2001, \mn@doi [\aap] {10.1051/0004-6361:20011308}, \href
  {http://adsabs.harvard.edu/abs/2001A%26A...379..279Q} {379, 279}

\bibitem[\protect\citeauthoryear{{Santos} et~al.,}{{Santos}
  et~al.}{2017}]{santos2017}
{Santos} N.~C.,  et~al., 2017, \mn@doi [\aap] {10.1051/0004-6361/201730761},
  \href {http://adsabs.harvard.edu/abs/2017A%26A...603A..30S} {603, A30}

\bibitem[\protect\citeauthoryear{{Sestovic}, {Demory}  \& {Queloz}}{{Sestovic}
  et~al.}{2018}]{Sestovic2018}
{Sestovic} M.,  {Demory} B.-O.,   {Queloz} D.,  2018, preprint, \href
  {http://adsabs.harvard.edu/abs/2018arXiv180403075S} {} (\mn@eprint {arXiv}
  {1804.03075})

\bibitem[\protect\citeauthoryear{{Sozzetti}, {Torres}, {Charbonneau}, {Latham},
  {Holman}, {Winn}, {Laird}  \& {O'Donovan}}{{Sozzetti}
  et~al.}{2007}]{sozzetti:2007}
{Sozzetti} A.,  {Torres} G.,  {Charbonneau} D.,  {Latham} D.~W.,  {Holman}
  M.~J.,  {Winn} J.~N.,  {Laird} J.~B.,   {O'Donovan} F.~T.,  2007, \mn@doi
  [\apj] {10.1086/519214}, \href
  {http://adsabs.harvard.edu/abs/2007ApJ...664.1190S} {664, 1190}

\bibitem[\protect\citeauthoryear{{Toner} \& {Gray}}{{Toner} \&
  {Gray}}{1988}]{toner1988}
{Toner} C.~G.,  {Gray} D.~F.,  1988, \mn@doi [\apj] {10.1086/166893}, \href
  {http://adsabs.harvard.edu/abs/1988ApJ...334.1008T} {334, 1008}

\bibitem[\protect\citeauthoryear{{Vanderburg} et~al.,}{{Vanderburg}
  et~al.}{2015}]{Vanderburg2015}
{Vanderburg} A.,  et~al., 2015, \mn@doi [\apj] {10.1088/0004-637X/800/1/59},
  \href {http://adsabs.harvard.edu/abs/2015ApJ...800...59V} {800, 59}

\bibitem[\protect\citeauthoryear{{Weiss} et~al.,}{{Weiss}
  et~al.}{2013}]{Weiss2013}
{Weiss} L.~M.,  et~al., 2013, \mn@doi [\apj] {10.1088/0004-637X/768/1/14},
  \href {http://adsabs.harvard.edu/abs/2013ApJ...768...14W} {768, 14}

\bibitem[\protect\citeauthoryear{{West} et~al.,}{{West}
  et~al.}{2016}]{west2016}
{West} R.~G.,  et~al., 2016, \mn@doi [\aap] {10.1051/0004-6361/201527276},
  \href {http://adsabs.harvard.edu/abs/2016A%26A...585A.126W} {585, A126}

\makeatother
\end{thebibliography}




\appendix

\section{Radial Velocities}

\begin{table}
\centering
\caption{CORALIE Radial Velocities of EPIC229426032.}\label{tab:rvs_coralie}
\begin{tabular}{lrrrr}
\hline\hline
\multicolumn{1}{l}{BJD}& \multicolumn{1}{c}{RV}&\multicolumn{1}{c}{$\sigma$ RV}& \multicolumn{1}{c}{BIS}& \multicolumn{1}{c}{S}\\ 
\multicolumn{1}{l}{ ( -2450000)}& \multicolumn{1}{c}{(\mpers)}&\multicolumn{1}{c}{(\mpers)}& \multicolumn{1}{c}{(\mpers)}& \multicolumn{1}{c}{(dex)}\\ \hline
7942.60097&  $-22339.6$&  38.8&    $-78$&  0.2823  \\
7943.55647&  $-21951.7$&  38.2&    $-69$&  0.1812  \\
7943.57794&  $-22000.3$&  38.6&   $-157$&  0.2420  \\
7944.56592&  $-22378.4$&  38.6&    $-37$&  0.1935  \\
7944.58780&  $-22413.6$&  38.6&    $-23$&  0.2483  \\
7945.57852&  $-22204.1$&  38.5&    $-34$&  0.6665  \\
7945.60178&  $-22196.3$&  38.5&    $-55$&  0.2057  \\
7946.54859&  $-22423.8$&  38.5&    $-49$&  0.1501  \\
7946.56573&  $-22404.5$&  38.6&    $-75$&  0.2128  \\
\hline\hline
\end{tabular}
\end{table}

\begin{table}
\centering
\caption{HARPS Radial Velocities of EPIC229426032.}\label{tab:rvs_harps}
\begin{tabular}{lrrrr}
\hline\hline
\multicolumn{1}{l}{BJD}& \multicolumn{1}{c}{RV}&\multicolumn{1}{c}{$\sigma$ RV}& \multicolumn{1}{c}{BIS}& \multicolumn{1}{c}{S}\\ 
\multicolumn{1}{l}{ ( -2450000)}& \multicolumn{1}{c}{(\mpers)}& \multicolumn{1}{c}{(\mpers)}& \multicolumn{1}{c}{(\mpers)}& \multicolumn{1}{c}{(dex)}\\ \hline
8036.55780& $-22434.7$&  26.0&    $-11$ &0.2714 \\
8037.51703& $-22027.7$&  27.9&    $ 10$ &0.2652 \\
8038.51742& $-22472.5$&  32.1&   $-136$ &0.2711 \\
8039.53105& $-22085.3$&  24.3&    $-20$ &0.2252 \\
\hline\hline
\end{tabular}
\end{table}

\begin{table}
\caption{FEROS Radial Velocities of EPIC246067459.}\label{tab:rvs_feros}
\centering
\begin{tabular}{lrrr}
\hline\hline
\multicolumn{1}{l}{BJD}& \multicolumn{1}{c}{RV}&\multicolumn{1}{c}{$\sigma$ RV}& \multicolumn{1}{c}{BIS}\\ 
\multicolumn{1}{l}{ ( -2450000)}& \multicolumn{1}{c}{(\mpers)}&\multicolumn{1}{c}{(\mpers)}& \multicolumn{1}{c}{(\mpers)}\\ \hline
8062.63804& 8247.9& 16.1&  $-15.2$\\
8063.63606& 8415.9& 24.1&  $191.2$\\
8064.55872& 8193.9& 15.5& $-102.4$\\
8065.59564&	8173.3&	13.0&   $15.0$\\
8065.64809& 8199.5& 15.9&  $-84.7$\\
8066.51427& 8366.3& 14.5& $-144.5$\\
8109.53786&	8198.1&	16.5&   $71.0$\\
8110.54097&	8275.7&	15.5&  $140.0$\\
8111.54702&	8352.8&	14.5&   $81.0$\\
8113.54191&	8204.4&	15.5&   $-2.0$\\
8114.54353&	8334.6&	14.8&   $88.0$\\
\hline\hline
\end{tabular}
\end{table}


\bsp	
\label{lastpage}
\end{document}